%% file: main.tex
\documentclass[twocolumn, sort&compress, 5p]{elsarticle}
\input{initialisation.tex}

\journal{}
\date{}

\begin{document}
\definecolor{correccolorone}{rgb}{0.8, 0.33, 0.1}
\definecolor{correccolortwo}{rgb}{0.85, 0.0, 0.1}
\definecolor{deleted}{rgb}{0.66, 0.66, 0.66}
\definecolor{brickred}{rgb}{0.8, 0.25, 0.33}

\begin{frontmatter}

\title{A gravity-independent powder-based additive manufacturing process\\ tailored for space applications}

\author[1]{Olfa D'Angelo\corref{c1}}
\cortext[c1]{Corresponding author}
\ead{olfa.dangelo@mail.com}

\author[1,2]{Felix Kuthe}
\author[1,3]{Szu-Jia Liu}
\author[4]{Raphael Wiedey}
\author[5]{Joe M.~Bennett}
\author[6]{Martina Meisnar}
\author[6]{Andrew Barnes}
\author[7,1]{W.~Till Kranz}
\author[1,8]{Thomas Voigtmann}
\author[1]{Andreas Meyer}

\affiliation[1]{organization={Institut f\"ur Materialphysik im Weltraum, Deutsches Zentrum f\"ur Luft- und Raumfahrt (DLR)},
            addressline={Linder H\"ohe}, 
            city={K\"oln},
            postcode={51170}, 
            country={Germany}}

\affiliation[2]{organization={Labor f\"ur Regelungstechnik und Mechatronik, Institut f\"ur Produktentwicklung und Konstruktionstechnik, Technische Hochschule K\"oln},
            addressline={Betzdorfer Stra\ss{}e 2}, 
            city={K\"oln},
            postcode={50679}, 
            country={Germany}}
            
\affiliation[inst2]{organization={Department of Materials Science and Engineering, University of Toronto},
            city={Toronto},
            postcode={ON M5S 3E4}, 
            country={Canada}}

\affiliation[4]{organization={Institute of Pharmaceutics and Biopharmaceutics, Heinrich Heine University},
            addressline={Universit\"atsstra\ss{}e 1}, 
            city={D\"usseldorf},
            postcode={40225}, 
            country={Germany}}
            
\affiliation[5]{organization={STFC-UKRI, Rutherford Appleton Laboratory},
            city={Didcot},
            postcode={OX110QX}, 
            country={United Kingdom}}

\affiliation[6]{organization={ESA-RAL Advanced Manufacturing Laboratory, European Space Agency, ECSAT},            
			addressline={Fermi Avenue}, 
            city={Didcot},
            postcode={OX110FD}, 
            country={United Kingdom}}

\affiliation[7]{organization={Institut f\"ur Theoretische Physik},
            addressline={Universit\"at zu K\"oln}, 
            city={K\"oln},
            postcode={50937}, 
            country={Germany}}
           
\affiliation[8]{organization={Department of Physics, Heinrich-Heine-Universit\"at D\"usseldorf},
            addressline={Universit\"atsstra\ss{}e 1}, 
            city={D\"usseldorf},
            postcode={40225}, 
            country={Germany}}

\begin{abstract}
The future of space exploration missions will rely on technologies increasing their endurance and self-sufficiency, including manufacturing objects on-demand.
We propose a process for handling and additively manufacturing powders
that functions independently of the gravitational environment and with no restriction on feedstock powder flowability.
Based on a specific sequence of boundary loads applied to the granular packing,
powder is transported to the printing zone, homogenized and put under compression to increase the density of the final part. 
The powder deposition process is validated by
simulations that show the homogeneity and density of deposition to be insensitive to gravity and cohesion forces within a \acrfull{dem} model. We further provide
an experimental proof of concept of the process by
successfully \acrshort{3d} printing parts on-ground and on parabolic flight in weightlessness.
Powders exhibiting high and low flowability are used as model feedstock material
to demonstrate the versatility of the process,
opening the way for additive manufacturing of recycled material.
\end{abstract}


\glsresetall

\end{frontmatter}

\input{introduction.tex}

\section{Additive manufacturing method}\label{sec:process}
\input{AMmethod.tex}

\section{Simulation of powder flow}\label{sec:simulation}
\input{simulation.tex}

\section{Experimental proof of concept}\label{sec:experiment}
\input{Experiment.tex}

\section{Results \& discussion}\label{sec:results}
\input{Results.tex}

\section{Conclusion and outlook}\label{sec:conc}

To be used for in-space applications, 
\gls{am} methods will need to evolve into more versatile technologies with
increased reliability~\cite{Trujillo2017, DebRoy2018}.
We proposed an \gls{am} process
that 
enables part production
from powder independently of the gravitational environment and of material flowability. 
This process emancipates from the current limitations on granular feedstock, as it does not rely on highly flowable powder for material deposition.
Besides, it places no further geometrical constraints compared to ground-based \gls{pbf},
while remaining superior to extrusion-based processes by
allowing a wider range of materials.
Tested through \gls{dem} simulation and on parabolic flights, 
a proof of concept was provided for \gls{1g} and \gls{mug},
using as feedstock material 
a \SI{80}{\micro\meter}-diameter \gls{ps} powder modified to obtain a good flowability powder (smooth surface, \gls{ss}) and a low flowability powder (rough surface, \gls{rs}).
The \gls{rs} powder was shown to have lower flowability than the typical \gls{3d} printing metal powder Ti-64.
Analysis of samples that were \gls{3d} printed under gravity and in absence thereof shows that the powder
deposition is realized equally efficiently under both $g$-levels.
High reproducibility is found between samples manufactured from the same base-material in different gravitational environments, with a homogeneous pore size distribution and isotropy of the samples. The ground-printed samples show slightly higher anisotropy,
which we attribute to a layer-wise crystallization in our highly monodisperse powders.
A difference between samples realized with each base materials persists, 
which is attributed to the lower ability to pack densely for \gls{rs} powder.
Delamination is observed in none of the \gls{3d} printed samples.

All the samples analyzed in-bulk (\gls{ss} and \gls{rs}, under \gls{1g} and \gls{mug})
were manufactured using the same printing parameters.
The mild differences obtained in material deposition, show that the use of 
material with different flowability
is reflected in the process; yet all powders could be deposited and \gls{3d} printed.
This suggests
that the \gls{rs} powder could be more densely packed by using better suited printing parameters.
In particular, while the deposition process might be ineffective in increasing packing density if that is a material-dependent variable~\cite{Mari2014}, 
this difference could be amended \emph{during} consolidation. 
Ideally, the compression ratio could be increased as the consolidation is taking place, by maintaining a constant vertical pressure, triggering degassing and thereby increasing part's density, akin to hot isostatic pressing.
This simple amelioration would allow
for the drastic reduction in the
samples' porosity, and in turn
enable the control of
printed parts density (including density gradients throughout the additively manufactured part).

The possibility to compress the material layer after deposition also has a 
specific drawback: 
it implies that the material of the solidification window and the printing material are in contact before and most importantly during consolidation.
For the polymer powders used as model substances in the work presented here, 
the quartz glass plate was not reactive at the sintering temperature of \acrlong{ps} ($\sim \SI{200}{\celsius}$). 
However, to \gls{3d} print metal or ceramic powders would require significant modifications of the system:
if consolidation happens at higher temperature, 
material exchange between the molten powder and the glass plate
could result in the newly solidified layer sticking irreversibly to the solidification window. 
The production of metal and ceramic parts would hence require not only to adapt the energy source to provide accordingly higher energy,
but also a new study of affinity between the base-material and the window's material.
The material of the solidification window would have to be chosen accordingly, 
to minimize material exchange and avoid the powder layer remaining stuck to the window.
For example, 
the stability of a sapphire glass plate 
should be investigated when exposed repeatedly to
molten metal alloys.
Ensuring that no heat-induced chemical reaction happens 
will be the challenge to adapt this \gls{am} process to 
metal and ceramic powders.

Besides 
controlling parts' density by compressing the newly deposited powder layer, 
defect appearance is mitigated by
constant \textit{in-situ} monitoring during powder deposition.
The next step
will be to automatize the control loops (torque sensing during powder transport and image analysis during layer homogenization).
On the one hand, homogenization time could be optimized through the image analysis procedure proposed,
to minimize fabrication time.
On the other hand,
assessment of print quality and live-correction during manufacturing 
will allow \gls{am} to access a wider range of applications
by increasing stability in prints quality,
including application for space and \gls{ism}.

Using those strategies to ensure 
high printing quality 
without requirement on the rheology of the feedstock material
will facilitate the use of recycled materials for \gls{am}.
First tests on-ground have shown
that powder produced by closed-loop recycling (i.e.~by grinding former \gls{3d} printed parts)
can be directly used as raw material in the process presented here~\cite{Lopez2021thesis}. 
The possibility to not only reuse material from previous batches,
but also recycle former objects into new feedstock,
would drastically reduce costs associated with \gls{am}, on Earth as well as in space.

Focusing on the powder handling aspect,
the powder deposition process
was developed to allow for the use of powders regardless of their physical and rheological properties,
meaning that it functions for \emph{any} base-material (polymers, metals, ceramics\textellipsis).
The powder handling method 
could hence be adapted for granular transport even beyond \gls{3d} printing
in reduced gravity environments,
regardless of the material's flowability.
Notably, on sand-covered planetary surface (e.g.~the Moon, Mars or certain asteroids),
powder handling technologies will be necessary to process regolith, the main \textit{in-situ} resource
and a powder of notoriously poor flowability~\cite{Walton2007}.

As technological progress and space explorations will go hand-in-hand in the coming years,
the authors hope that
the \gls{am} process presented 
will be part of a movement to spur the development of \acrlong{ism} in general,
ultimately enabling long-term human presence in space.

\section*{Author contributions}

\textbf{Olfa D'Angelo}: conceptualization, investigation, methodology, data curation, and formal analysis, funding acquisition and project administration, writing -- original draft.
\textbf{Felix Kuthe}: investigation, methodology, data curation.
\textbf{Szu-Jia Liu}: investigation, data curation.
\textbf{Raphael Wiedey}: resources.
\textbf{Joe M. Bennett}: resources.
\textbf{Martina Meisnar}: resources.
\textbf{Andrew Barnes}: resources.
\textbf{W. Till Kranz}: supervision, writing -- review \& editing.
\textbf{Thomas Voigtmann}: methodology, data curation, validation, supervision, writing -- review \& editing.
\textbf{Andreas Meyer}: funding acquisition, project administration, resources, supervision, writing -- review \&
editing.

\section*{Supplementary material}
Three videos supplement this article.
\begin{itemize}[label={-}]
	\item Video 1: visualization of the \acrfull{dem} simulation results. The external container is not shown for the granulate to be visible. The simulation is run for a gravitational acceleration $g = 0$, with 76~000 particles of diameter $d = \SI{2}{\mm}$. In the video presented, the cohesion parameter $\kappa = 10^5$~\si{\newton\per\meter\square}. The color scale represents the magnitude of the speed at which the particles are moving, hence highlighting the particles in motion versus the static particles for each step fo the deposition process.
	\item Video 2: video from \textit{in-situ} monitoring by image capture from below the solidification window, in one of the apparatus used for experimental campaigns. Shown is the homogenization during one parabola of \SI{22}{\second} (hence fully in microgravity), for \gls{rs} powder, corresponding to the $4^{\text{th}}$ layer deposited during the third campaign day.
\end{itemize}

\section*{Acknowledgments}
\noindent
ODA gratefully acknowledges financial support from 
\acrshort{esa} \acrshort{eac} through the \acrshort{npi} contract 
4000122340 on \enquote{Physical Properties of Powder-Based \acrshort{3d}-Printing in Space and On-Ground} supported by Aidan Cowley,
and of the \acrshort{dlr}/\acrshort{daad} Research Fellowship
91647576,
as well as the \acrshort{esa} Education Fly Your Thesis!~2019 flight opportunity and sponsoring through the GrainPower project, in particular Nigel Savage.
The entire GrainPower team is also acknowledged:  Merve Se\c{c}kin, Abeba Birhane and Tolga Bast\"urk, thank you.
Many thanks to the Novespace team for their friendly assistance, and in particular Thomas Villatte.
ODA also acknowledges Fanny Schaepelynck for her help with the schematic in Fig.~\ref{fig:simplified_schematic_art},
and the
\acrshort{esa}-\acrshort{ral} Advanced Manufacturing Laboratory for making the \acrshort{ft4} rheological measurements possible.
WTK acknowledges funding from the DFG through grant number KR4867/2.
ODA and SJL acknowledge support from the DAAD RISE 2019 program under the grant number 57467143 for the project \enquote{Characterizing powder flow in a prototype microgravity powder-based \acrshort{3d} printer}.

\bibliographystyle{elsarticle-num} 
\bibliography{bibliography.bib}{}

\end{document}

%% file: initialisation.tex
\usepackage[hidelinks]{hyperref}
\usepackage{graphicx}      
\usepackage{amsmath}      

\usepackage{bm}
\renewcommand{\vec}[1]{\text{\bfseries #1}}

\usepackage{graphicx}
\usepackage{dcolumn}
\usepackage{bm}
\usepackage{tikz}
\usetikzlibrary{shapes.geometric, arrows}
\usetikzlibrary{positioning}
\usetikzlibrary{shapes,arrows}
\usetikzlibrary{intersections}
\usepackage{csvsimple}
\usepackage{changepage}
\usepackage[tbtags]{mathtools}
\usepackage{siunitx}
\usepackage{csquotes}
\usepackage{enumerate}
\usepackage{enumitem}
\usepackage{float}
\usepackage{subfigure}
\usepackage[
	nonumberlist, 				
	acronym,      				
	nomain,						
	nopostdot,					
]{glossaries}
\usepackage{glossary-superragged}
\usepackage{color, colortbl}

\usepackage{wrapfig}
\usepackage{natbib}
\usepackage{pgfplots}
\usetikzlibrary{arrows}
\usepackage{amsmath}
\pgfplotsset{compat=newest}
\usepgfplotslibrary{fillbetween}
\usetikzlibrary{calc}
\def\centerarc[#1](#2)(#3:#4:#5)
{ \draw[#1] ($(#2)+({#5*cos(#3)},{#5*sin(#3)})$) arc (#3:#4:#5); }

\usetikzlibrary{external}
\usepackage{amssymb}
\let\vec\mathbf
\usepackage{nicefrac}

\usepackage{multirow}
\usepackage{tabularx}
\usepackage{booktabs}
\usepackage{array}
\usepackage{longtable}
\newcolumntype{L}[1]{>{\raggedright\let\newline\\\arraybackslash\hspace{0pt}}m{#1}}
\newcolumntype{C}[1]{>{\centering\let\newline\\\arraybackslash\hspace{0pt}}m{#1}}
\newcolumntype{R}[1]{>{\raggedleft\let\newline\\\arraybackslash\hspace{0pt}}m{#1}}

\usepackage[normalem]{ulem}

\newacronym{3d}{3D}{three dimensional}
\newacronym{am}{AM}{additive manufacturing}
\newacronym{fdm}{FDM}{fused deposition modeling}
\newacronym{ism}{ISM}{in-space manufacturing}
\newacronym{iss}{ISS}{International Space Station}
\newacronym{fcb}{FCB}{Functional Cargo Block}
\newacronym{dem}{DEM}{discrete element method}
\newacronym{md}{MD}{molecular dynamics}
\newacronym{dc}{DC}{direct-current}
\newacronym[plural=PFCs,firstplural=parabolic flight campaigns (PFCs)]{pfc}{PFC}{parabolic flight campaign}
\newacronym{fft}{FFT}{Fast Fourrier Transform}
\newacronym{cad}{CAD}{Computer Assisted Design}
\newacronym{ptfe}{PTFE}{polytetrafluoroethylene}
\newacronym{ps}{PS}{polystyrene}
\newacronym{nasa}{NASA}{National Aeronautics and Space Administration}
\newacronym{esamm}{ESAMM}{Extended Structure Additive Manufacturing Machine}
\newacronym{amf}{AMF}{Additive Manufacturing Facility}
\newacronym{us}{US}{United States}
\newacronym{usa}{USA}{United States of America}
\newacronym{bmgs}{BMGs}{Bulk Metallic Glasses}
\newacronym{esa}{ESA}{European Space Agency}
\newacronym{si}{SI}{International System of Units, abbreviated from French \textit{Syst\`{e}me International (d'unit\'{e}s)}}
\newacronym{dlr}{DLR}{German Aerospace Center} 
\newacronym{liggghts}{LIGGGHTS}{\acrshort{lammps} Improved for General Granular and Granular Heat Transfer Simulations}
\newacronym{lammps}{LAMMPS}{Large-scale Atomic/Molecular Massively Parallel Simulator}
\newacronym{sjkr}{SJKR}{Simplified Johnson-Kendall-Roberts}
\newacronym{ded}{DED}{Directed Energy Deposition}
\newacronym{slm}{SLM}{Selective Laser Melting}
\newacronym{sls}{SLS}{Selective Laser Sintering}
\newacronym{eva}{EVA}{Extra-Vehicular Activity}
\newacronym{sem}{SEM}{Scanning Electron Microscopy}
\newacronym{RPM}{RPM}{Ramdom Positioning Machine}
\newacronym{rpm}{rpm}{revolutions per minute}
\newacronym{rise}{RISE}{Research Internships in Science and Engineering}
\newacronym{daad}{DAAD}{German Academic Exchange Service, abbreviated from German \textit{Deutscher Akademischer Austauschdienst}}
\newacronym{fsm}{FSM}{finite-state machine}
\newacronym{ir}{IR}{infrared}
\newacronym{pcbs}{PCBs}{Printed Circuit Boards}
\newacronym{pcb}{PCB}{Printed Circuit Board}
\newacronym{mcr}{MCR}{Modular Compact Rheometer}
\newacronym{sff}{SFF}{Solid Freeform Fabrication}
\newacronym{uv}{UV}{ultraviolet}
\newacronym{abs}{ABS}{acrylonitrile butadiene styrene}
\newacronym{hpde}{HPDE}{high density polyethylene}
\newacronym{pei}{PEI}{polyetherimide}
\newacronym{bff}{BFF}{BioFabrication Facility}
\newacronym{lens}{LENS}{Laser Engineered Net Shaping}
\newacronym{cnc}{CNC}{Computer Numerical Control}
\newacronym{ebf3}{EBF$^3$}{Electron Beam Free-Form Fabrication}
\newacronym{leo}{LEO}{Low Earth Orbit}
\newacronym{pc}{PC}{polycarbonate}
\newacronym{crissp}{CRISSP}{Customisable Recyclable International Space Station Packaging}
\newacronym{Athena}{Athena}{Advanced Telescope for High-ENergy Astrophysics}
\newacronym{lbm}{LBM}{Laser Beam Melting}
\newacronym{bam}{BAM}{Federal Institute for Materials Research and Testing, abbreviated from German \textit{Bundesanstalt f\"{u}r Materialforschung und-pr\"{u}fung}}
\newacronym{pbf}{PBF}{powder bed fusion}
\newacronym{eb}{EB}{Electron Beam}
\newacronym{2d}{2D}{two dimensional}
\newacronym{ft4}{FT4}{Freeman Technology 4 Powder Rheometer}
\newacronym{dsc}{DSC}{Differential Scanning Calorimetry}
\newacronym{pmma}{PMMA}{polymethylmethacrylate}
\newacronym{1g}{$1g$}{on-ground}
\newacronym{mug}{$\mu g$}{weightlessness}
\newacronym{bcm}{BCM}{Box Counting Method}
\newacronym{mct}{MCT}{Mode Coupling Theory}
\newacronym{gmct}{gMCT}{granular Mode Coupling Theory}
\newacronym{itt}{ITT}{Integration Through Transients}
\newacronym{mfc}{MFC}{Mass Flow Controller}
\newacronym{ct}{CT}{computed tomography}
\newacronym{xct}{XCT}{X-ray computed tomography}
\newacronym{cv}{CV}{curriculum vitae}
\newacronym{pi}{PI}{principal investigator}
\newacronym{osp}{OSP}{orthogonal superimposed perturbation}
\newacronym{npi}{NPI}{Network Partnering Initiative}
\newacronym{ecsat}{ECSAT}{European Centre for Space Applications and Telecommunications}
\newacronym{eac}{EAC}{European Astronaut Centre}
\newacronym{estec}{ESTEC}{European Space Research and Technology Centre}
\newacronym{fps}{fps}{frames per second}
\newacronym{pdf}{pdf}{probability density function}
\newacronym{al}{Al}{aluminium}
\newacronym{ss}{\textit{SS}}{\textit{Smooth Surface}}
\newacronym{rs}{\textit{RS}}{\textit{Rough Surface}}
\newacronym{rcp}{rcp}{random close packing}
\newacronym{ral}{RAL}{Rutherford Appleton Laboratory}
\newacronym{sac}{SAC}{surface area coverage}

%% file: introduction.tex
As human reach into space expands, need arises for machines that work under extreme conditions -- notably, in absence of gravity. 
Space exploration missions are severely constrained by payload capacity, and relying upon ground-support would largely increase the risk of failure of such mission~\cite{Skomorohov2016}.
As long endurance missions must be able to solve unexpected problems autonomously,
a sustainable approach is the only valid alternative for human spaceflight to non-low Earth orbit: missions' self-reliability will be a key to their success~\cite{Owens2017}.

A vision for space exploration is \gls{ism}: fabrication, assembly and integration of small to large structures directly in space~\cite{3Dp_in_space_book, Skomorohov2016}. 
\gls{ism} has the potential to significantly enhance the self-sustainability of missions, as it could support space exploration missions by maintenance, repair and production of objects without depending on ground-support~\cite{Prater2019IAC}. 
Having autonomous manufacturing capabilities in space also opens the possibility to adapt the design of structural systems to their final function in zero-gravity environment, instead of over-engineering them to resist terrestrial gravity and launch. Approximately 30\% of the structural mass of payload shipped to space today could be saved if the launch load constraints could be avoided~\cite{Trujillo2017}, 
representing high economical and ecological gains. 

\Gls{am}, also known as
\gls{3d} printing,
encompasses technologies that have two essential advantages for space applications:
first, compared to subtractive technologies, they reduce the quantity of waste material produced \cite{Owens2017}.
Second, they open the possibility to access virtually any geometry, 
rendering obsolete the geometrical constraints of classical manufacturing techniques.
The possibility to recycle 
former objects into new feedstock material 
would 
optimize payload all the more by up-cycling waste to
minimize the necessary raw material mass.

Strictly speaking of manufacturing, 
\gls{am} already is a permanent tool in space: extrusion-based \gls{3d} printers have been on-board the \gls{iss} since 2014~\cite{Prater2016, Prater2018}. 
This so-called \gls{amf} has 
produced over 200 parts in orbit to this day, including spare parts and tools~\cite{Prater2018},
highlighting \gls{am} as an essential tool for future space missions.
However, extrusion-based technologies suffer inherent limitations.
First, they are restricted to
materials showing continuous viscosity decrease with increasing temperature, which 
makes such technique most adapted to thermoplastics \cite{Gibson2018}.
Moreover, specially manufactured filament feedstock is necessary, which has to be carried along at the cost of large storage volume.
Besides, filament-based technologies have limited resolution, restrained by the diameter of the deposited filament; parts produced are typically prone to delamination and highly anisotropic in their mechanical and physical properties~\cite{Goh2020}. 

As on Earth, different manufacturing technologies should be available for space in order to respond to the variety of needs.
Among \gls{am} technologies available on-ground,
\gls{pbf} technologies offer the highest resolution~\cite{DebRoy2018} and most versatile techniques~\cite{Schmid2016, Yap2015, Sing2017}.
The difficulty to handle powders in reduced gravity~\cite{Blum2000, Weidling2012, Love2014, Wilkinson2005, Lopez2021}
has hitherto been an obstacle to further development of powder-based technologies for \gls{ism}.
A recent breakthrough showed the possibility to adapt \gls{slm} of metal powders to \gls{mug}.
The method proposed by Zocca et al.~\cite{Guenster2017, Zocca2014, Zocca2019} consists of stabilizing the powder bed by 
applying a pressure difference between the bottom and the top of the powder-bed using a suction pump.
Tested in \gls{mug} between 2017 and 2019, it enabled {the production of} parts from ceramic and stainless steel powders~\cite{Zocca2019} while depositing the powder in weightlessness {during \glspl{pfc}}.
Despite the tremendous achievement
of producing the first 
parts manufactured from powder deposited in \gls{mug},
this method suffers specific drawbacks,
detailed by Zocca et al.~\cite{Zocca2019}:
as large closed surfaces would prevent the air flow from going through the parts and accessing the next deposited powder layer,
closed horizontal surfaces cannot be printed. Using open structures connected by vertical walls, the 
thickness of those walls is limited to approx.~\SI{2}{\mm}.
Moreover,
the required pump power increases with the powder bed height, 
necessitating
a large quantity of hardware.
Finally, powders which include many fines cannot be processed because the filling of interstitial volume becomes too high and annihilates the effect of the air flow.
It is also noteworthy that as for all powder-based \gls{am} processes used on-ground,
the powder deposition step is based on the high {flowability}
of the powder feedstock~\cite{Vock2019, Seyda2017,Kiani2020}.
This implies strict requirements on the manufacturing and storage of the powder,
difficult to provide in remote, extreme environments.
Furthermore, it complicates direct re-usage of material from previous batches and prohibits closed-loop recycling.
Such inherent drawbacks
question the superiority of additive technologies for \gls{ism},
as the limitation to neither reuse nor recycle powder 
amounts to the production of large quantities of waste material.

The contours of a technological gap appear:
to be suitable for space applications, an \gls{am} technology would 
combine the assets of \gls{pbf} with the possibility to use powders regardless of their flow-properties,
and be robust against changes in $g$-level.
While powder handling remains an important issue \gls{1g}~\cite{Jaeger1996,Forterre.2008},
and in absence of constitutive equations enabling large-scale predictions of granular flows in any environmental conditions~\cite{Aranson2006},
powder handling technologies for space applications face specific challenges.
Primarily, to fulfill gravity-independence, the body force created on each particle by gravity cannot be used as transport mechanism, and
normal pressure applied on top of the granular packing 
cannot be used to induce powder flow,
since any such normal pressure is reoriented horizontally
by the granular packing according to the Janssen effect~\cite{Janssen1895}.
Furthermore, versatility in raw material is required to ease powder storage and recycling:
besides the higher stress required to overcome 
friction and mechanical locking between particles for low flowability powders, 
a jammed phase~\cite{OHern2003, Cates1999} also appears at lower packing density for particles showing angular shape and rough surface state~\cite{Mari2014}.
The appearance of a jammed region in a larger packing is a challenge in powder handling,
and can draw complete industrial processes to a halt.
Recent studies show that
by changing the force balance acting on each particle,
\gls{mug} also appears to decrease the \emph{rearrangeability} of the particles' spatial configuration~\cite{Lopez2021b},
ergo facilitating jamming.
Actively avoiding the appearance of a jammed phase hence becomes yet another requirement to ensure reliable functioning of powder handling and \gls{3d} printing for space applications. 

In the present work, we
propose a method to \gls{3d} print powders, regardless of the rheological properties of the feedstock raw material, and independently of the gravitational environment.
Our method rests on a mechanism for powder transport and homogenization, as well as solidification of the granular material, in a closed container. 
In the following, \emph{solidification} describes 
the consolidation process by which the granular material is transformed into a solid, coherent object,
by any appropriate physical process chosen depending on the powder at hand.
Robustness
against gravity-variations is achieved 
by depositing powder solely using 
driving mechanisms shown to induce similar response regardless of the gravitational environment -- namely, shear~\cite{Murdoch2013a} and shaking~\cite{Opsomer2012} of the granular material.
We focus on the aspects of flow properties posed by different powders and different gravitational environment, using two \gls{ps} demonstrator powders of different flowability on ground and in \gls{mug}. We demonstrate qualitatively through computer simulation, and quantitatively and directly through experiments performed on \glspl{pfc}, that the method is able to produce sintered parts of \gls{ps} powder that are dense and homogeneous. A key result of our work is that the proposed process is capable of handling powders also of poor flowability, and that the microscopic properties of the finally sintered part are nearly independent of the gravitational environment under which they have been produced.

After describing the \gls{am} method in Sec.~\ref{sec:process}, \gls{dem} simulation will be used in Sec.~\ref{sec:simulation} to model the powder handling process. An experimental implementation of the \gls{am} process will follow in Sec.~\ref{sec:experiment}, providing a proof of concept on-ground and in weightlessness through \glspl{pfc}. Parts manufactured from materials of variable flow-behaviors, under gravity conditions of \gls{1g} and \gls{mug}, will be analyzed in Sec.~\ref{sec:results}, 
enabling to assess the performances of the \gls{am} process.
Section~\ref{sec:conc} provides concluding remarks and and outlook.

%% file: AMmethod.tex
\Gls{am} generally amounts to multiple iterations of two main steps: material deposition, followed by material solidification.
In the case of \gls{pbf}, powder deposition 
consists in creating a thin and homogeneous layer of granular material,
which will then be selectively solidified -- the latter being achieved by melting, sintering, or the addition of an extrinsic phase.
The present powder-based \gls{am} method aims to conduct the material deposition step without relying on the gravitational environment, nor imposing constrains on the raw material flow-properties.

\subsection{Process confinement}
\begin{figure}
	\centering
	\includegraphics[width=\linewidth]{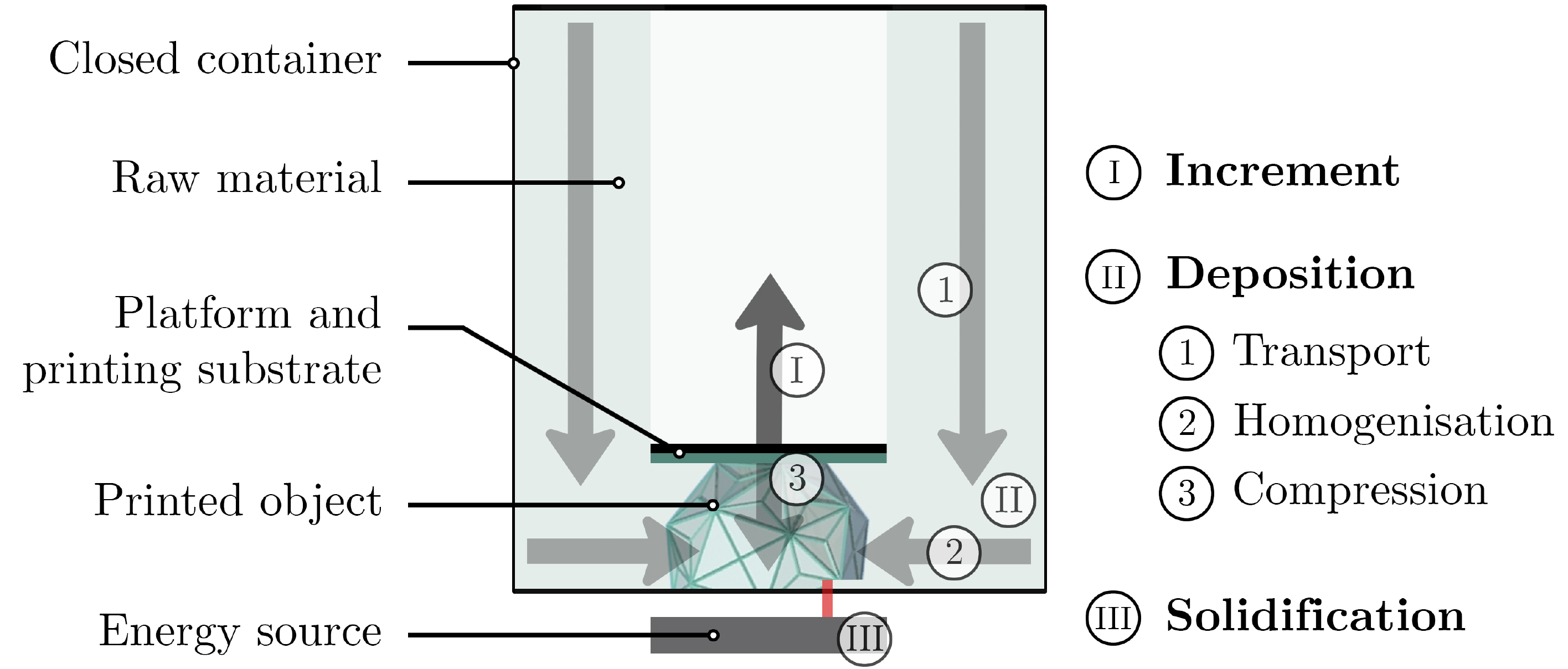}
	\caption{Principle of the \gls{am} method. The feedstock material is confined in a closed container, inside which the entire process takes place. The stepwise process consists of: I.~incremental platform rise, II.~powder deposition and III.~selective solidification of the newly deposited layer. The powder deposition step encompasses the following powder movements: 1.~vertical downward transport towards the bottom of the container, 2.~horizontal homogenization to create evenly distributed layers under the platform, and 3.~compression of the newly deposited layer.}
	\label{fig:super_simplified_schematic_method}
\end{figure}

The approach proposed here consists of confining the raw material in a closed container inside which the entire process takes place
(whereas in conventional \gls{pbf}, deposition and solidification happen on
an open powder bed~\cite{DebRoy2018,Zocca2019}). 
The deposition step amounts to controlling the powder flow inside the closed space to force material to the desired location.
Inside the container is a platform or printing substrate
 on which the object will be \gls{3d} printed upside down; the desired location for each new layer to be deposited is the horizontal space under this printing platform. At the beginning of the manufacturing process, the platform is placed at the bottom of the container. It then moves up in discrete steps, each iteration allowing one new layer to be \gls{3d} printed underneath the platform and portion of object already printed. Fig.~\ref{fig:super_simplified_schematic_method} illustrates this method.

Handling the confined raw material is achieved by moving the container itself 
to force the material to flow towards the desired location. The powder displacement can be divided into two types of movement: 1.~the vertical, downward powder transport and 2.~the horizontal, planar movement to create homogeneous layers under the platform.
Once deposited, the powder can be selectively solidified. The solidification also takes place inside the container: an energy input is provided from outside to the material inside the container through the bottom wall of the container, transparent to the type of energy used to solidify the raw material.

\subsection{Powder deposition}
\begin{figure}
	\centering
	\includegraphics[width=\linewidth]{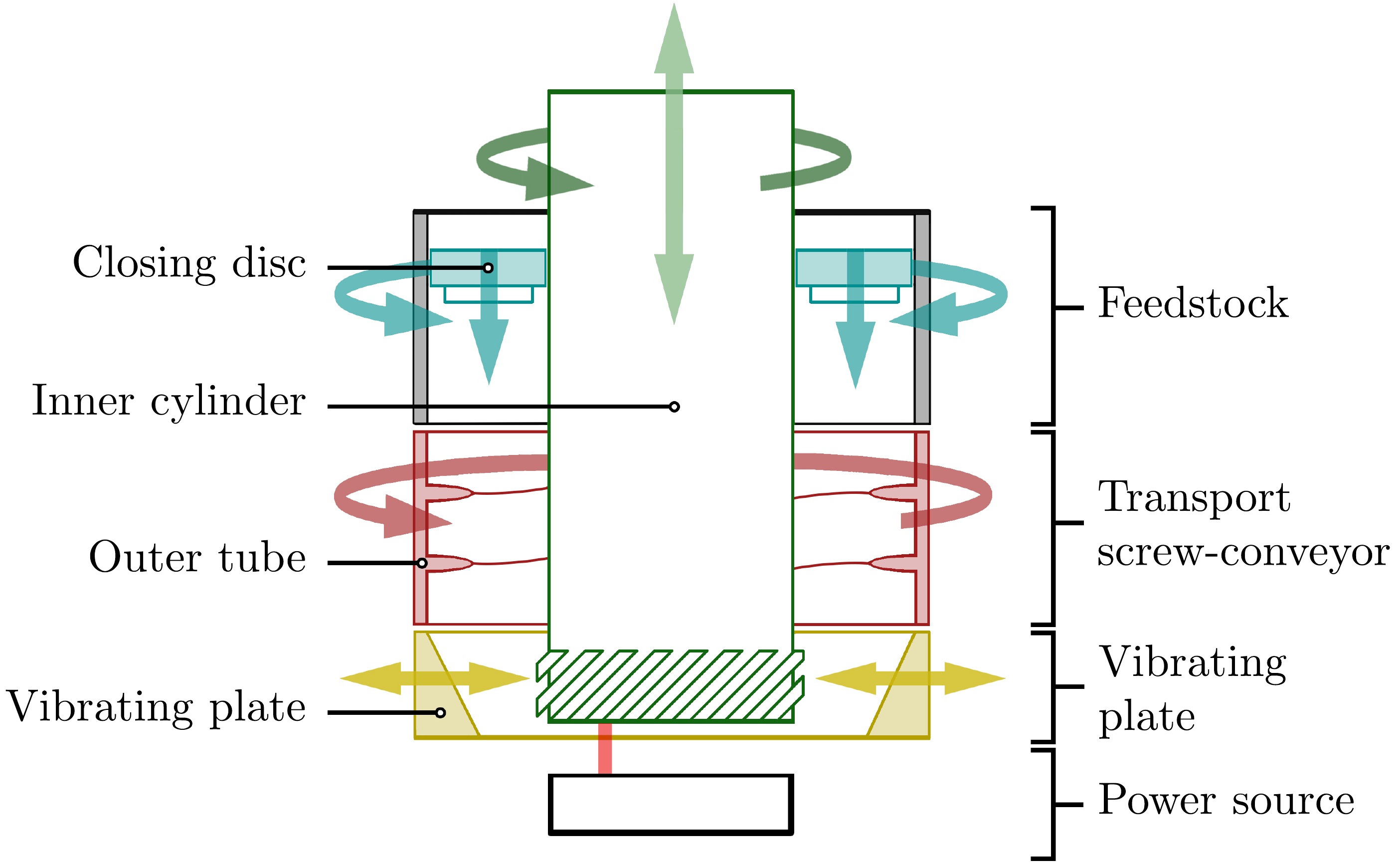}
	\caption{Schematic representation of the different stages of the \gls{am} process. The powder container is divided in four parts, each moving independently: the closing disc descends while describing oscillatory rotation; the inner {cylinder} rises stepwise, rotates and descends to compress the newly deposited layer; the outer tube rotates to activate the screw conveying system; the vibrating plate produces horizontal shaking. The apparatus can be divided in four stages (from top to bottom): the powder feedstock, the material transport by screw conveyor, the homogenization by horizontal vibration and the selective solidification. }
	\label{fig:simplified_schematic_art}
\end{figure}

Motion is imposed on the powder exclusively by movements of the container itself, i.e.~only through boundary forces. 
The container is cylindrical, axially symmetrical about the axis along which the platform rises.
The process is schematized in Fig.~\ref{fig:simplified_schematic_art}
through the motion of each part.
 
Since direct compression of the powder might lead to a fully jammed phase,
another transport mechanism must be sought. 
Shear stress applied to a granular packing creates a primary flow independent of the gravitational field~\cite{Murdoch2013, Murdoch2013a}, and
it can be applied to powders regardless of their flow-properties.
Therefore, shear will be the preferred mechanism to trigger controlled granular motion.
As shear can also lead to shear-jamming~\cite{Seto2019}, 
a superposition of shear forces in different spatial directions is used to avoid the creation of stable force chains and thus to preempt jamming.

Also in an effort to avoid putting the granular packing under purely normal compression,
the printing platform moving through the container is not a platform but the base of a cylinder; hence, no powder can remain compressed between the platform and the upper wall of the container as the platform moves upward to give space to the printed part. The rising cylinder on which the printing substrate is installed is labelled \emph{inner cylinder}.

The rise of the inner cylinder at each new iteration increases the volume available for the powder under the printing platform, but also in the container in general.
To maintain the total powder volume fraction constant throughout the process, 
the volume gain is compensated by lowering the part 
closing the container on its upper section. Labelled \emph{closing disc}, this annular shape links 
the outer wall of the container to the inner cylinder; it descends on the feedstock area, to push downward
the raw material stocked there.
Again to avoid normal compression, 
the closing disc describes oscillatory rotation {around the cylinder axis} while descending. It rotates alternatively in each direction at a frequency of \SI{1}{\hertz},
and it
is equipped with vanes penetrating the powder bed. The oscillatory motion forces the material to reorganize regularly, destroying and reforming the \enquote{fragile skeleton}~\cite{Cates1999} of force chains supporting the downward pressure. At each reorganization, the particles are pushed to a position of (temporary) stability lower along the $z$-axis than their previous one.
It is ensured in this manner that the material is periodically pushed downwards and enters the next step of the powder handling system: the transport area. 

Vertical transport of granular material in a closed container has been widely studied on-ground, for instance in the case of silo discharge, showing that normal pressure applied on top of a granular packing is reoriented horizontally~\cite{Janssen1895}.
The present application poses a supplementary requirement: the body force created on each particle by gravity can also not be used as transport mechanism, as it would render the powder handling method gravity-reliant. 
Therefore, a screw conveying system is used to transport the powder vertically:
the rotating outer container (labelled \emph{outer tube}) is equipped with helical blades that shear the material downward as they rotate. 
This mechanism enables the handling of a wide range of powders regardless of their physical or rheological properties, as will be demonstrated below.
 
\begin{figure}[h!]
	\centering
	\subfigure[~Section view]{
	\includegraphics[width=0.47\linewidth]{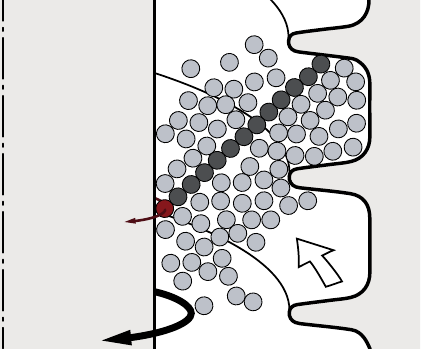}} \hfill
	\subfigure[~Top view]{
	\includegraphics[width=0.47\linewidth]{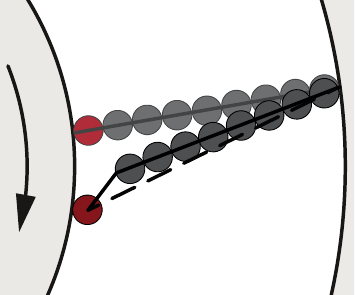}}
	\caption{Illustration of the superposition of force chains (in dark grey) imposed by the screw conveyor motion (white arrow), and motion created by rotation of the inner cylinder (black arrow), disrupting the end-components of those force chains (red particle).}
	\label{fig:superimposition}
\end{figure}

During granular shear, force chains form oblique to the direction of shear~\cite{Mair2007}. 
Force chains are the lines of force through dense granular packings; as such, a fragile networks of force chains is desired in order to transmit force from the boundary to the bulk.
If {the force chains however} percolate into a stable configuration,
they might transfer the load directly 
from the screw conveyor to the inner cylinder, creating a jammed phase that is stable against further motion.  
To ensure that force chains forming are intermittently destroyed, 
a secondary force field is superposed by rotating the inner {cylinder} simultaneously with the screw conveyor.
{The inner cylinder is equipped at its bottom with blades to enhance powder-powder contact during the rotation.}
The mechanism used to defuse the force chains is illustrated by Fig.~\ref{fig:superimposition};
superposition of perturbations in different directions have been used previously to tune jamming in dense shear thickening suspensions~\cite{Niu2020}.
In the present case,
the inner {cylinder}
imposes
a torque on the particles in contact with it as it rotates, which \enquote{elongates} the chain, thereby destabilizing it by rolling particles out of the main stress direction~\cite{Edwards1989, Cates1999}.
Superposing a secondary flow direction
forestalls jamming by defusing the long force chains as they appear, 
constantly imposing plastic deformation to the packing.
It is noteworthy, that if surface friction increases stability of the force chains, both mechanisms used in the superposition of directions of drive are enhanced by {an increase in particle-surface friction}, as {friction} also renders contacts between particles and container's surface more stable.

Once the powder particles have been brought to the bottom of the container, the powder needs to be spread homogeneously on the entire bottom surface. 
To do so, the inner cylinder rises by more than one layer-height, 
leaving under the printing platform an empty volume greater than that of a powder layer.
Then, homogenization is realized by applying horizontal shaking to the bottom of the container.
Granular homogenization by planar shaking is a well-know mechanism on-ground~\cite{Nadler2011, Khefif2018}. 
In \gls{mug}, 
shaking 
a confined granular sample leads to the formation of
a large cluster
bouncing around the middle position of the axis along which shaking is applied. 
This has been shown through
simulation~\cite{Opsomer2011, Opsomer2012, Noirhomme2015} 
and experiments~\cite{Kudrolli1997, Falcon1999, Opsomer2017}.
Using alternated shaking along $x$- and $z$-axis, 
powder is shifted towards the middle position.
Powder coming from the sides is added as it reaches the lower part of the container.
Shaking continues until powder completely fills the bottom layer. 
The duration needed to reach this state is 
monitored \textit{in-situ} to ensure that the entire printing surface is filled with powder (see Sec.~\ref{subsec:insitu}).

After having been transported down and homogeneously distributed horizontally, the newly deposited powder layer undergoes normal compression by the platform descending onto it.
The powder layer is hence compressed between the bottom wall of the container and the previously solidified layer sitting on top of the platform, increasing the packing density up to close packing.
The compression ratio, expressed as a function of the layer height,
is a printing parameter.

Finally, the newly deposited powder layer can be selectively solidified through the bottom wall of the powder container, by the energy source placed outside.
Since this wall must be transparent to the type of energy used to solidify the material, it is labelled \textit{solidification window}.

\subsection{\textit{In-situ} monitoring}\label{subsec:insitu}
Quality and repeatability have been identified as the Achille's heel of \gls{am}~\cite{Tapia2014}.
The problem of defects appearing in printed parts constitutes a major obstacle for \gls{am} in industrial applications, as
layer-wise material deposition increases the risk of defects appearance; yet it also enables 
a direct insight into the bulk of the object while it is manufactured.
Using this specificity for
\textit{in-situ} monitoring 
would enable to spot defects and hinder their appearance~\cite{Grasso2017, Everton2016}.

The present process is designed to allow for closed-loop control by \textit{in-situ} monitoring.
Primarily, 
the torque needed 
to rotate the inner cylinder mono-directionally during material transport is recorded.
The torque developed during oscillatory rotation 
of the closing disc
as it descends is also recorded, providing a second source of information on the raw powder's rheological behavior.
The adaptive control loop 
allows for the immediate reaction
to changes in flowability upon changes in environmental conditions.
It must be noted, that here flowability does not refer to an inherent property of the material, but to the flow exhibited by a powder in the given conditions and environment in which it is processed, which includes -- but is not limited to -- the gravitational environment.
Hence, the torque developed during powder transport is compared
to a scale established \textit{a priori},
giving the typical duration needed to deposit a material as a function of its flow response. 
This adaptive closed control loop allows
for the optimization of
powder deposition without being limited to situations formerly encountered.

In parallel, 
a quality assurance system is implemented to monitor the appearance of defects during material homogenization.
The solidification window is transparent not only to the solidification energy but also to visible light; hence, live imaging
captures the progression of the powder layer homogenization from below. 
This 
image-analysis procedure facilitates the detection and quantification of
heterogeneities in the powder layer,
continuing the material deposition procedure as long as the chosen metrics have not dropped under a threshold set to identify an acceptable degree of homogeneity.

\begin{figure}[h!]
    \centering
    \includegraphics[width=\linewidth]{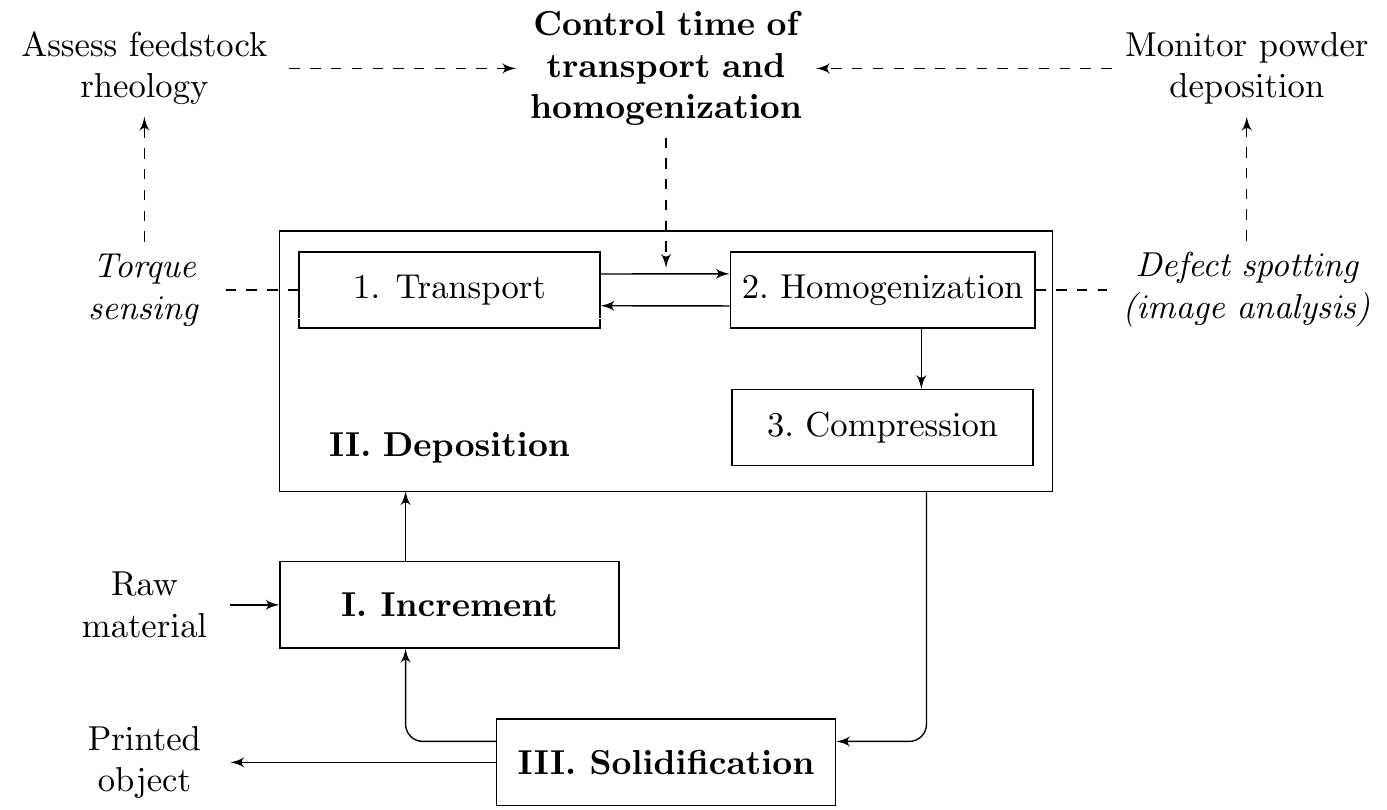} 
    \caption{Manufacturing procedure, including closed control loop used to optimize the duration of powder deposition and ensure defect-free powder layers.}
    \label{fig:DIAG_processmonitoring}
\end{figure}

Proposing
a step toward autonomous manufacturing,
concurrent use of the two \enquote{probe-and-adapt} systems
mentioned above (\textit{i.e., torque-based rheometrical feedback and \textit{in-situ} monitoring)} does
not only provide
traceability of defects, but their 
automated correction, ensuring 
constant and reliable manufacturing quality.
The full printing procedure, including \textit{in-situ} monitoring mechanisms, is schematized in Fig.~\ref{fig:DIAG_processmonitoring}.

This process was developed to function independently of 
feedstock flowability and variations in $g$-level.
To observe the effect of varying those parameters 
on the powder deposition efficiency,  
for the sake of comparison,
the process shall be tested in those different situations with the same (fixed) printing parameters. For this reason, the \textit{in-situ} probing is not automated in this series of experiments.
Printing parameters include powder deposition time, rotation speed of the different parts, sintering time and compression ratio. 
In an effort to limit the \gls{mug}-time necessary, 
the experimental campaign is preceded
by a preliminary simulation study.
Besides the optimization of rotation speeds (not shown here), 
the simulation aims to
minimize
\gls{mug}-time needed for empirical 
parameter screening,
by verifying 
if the same parameters can be used for manufacturing 
high and low flowability materials, in \gls{1g} and \gls{mug}.

%% file: simulation.tex
The following simulation study of material deposition is used to
validate the printing parameters prior to experiment, for both \gls{1g} and \gls{mug}.
Lacking precedent on which to rely on for comparison,
and in an effort to narrow the 
possible sources of variations,
the same printing parameters are used on all the situations presented.
The simulation study 
guides the choice of these printing parameters,
while
minimizing the risk of failure, notably for \gls{mug} experiments,
by testing the qualitative influence
of key parameters, such as gravity and increase in powder cohesion.

\subsection{Simulation methodology}

\Gls{dem} simulation~\cite{Luding2005} is used to validate
qualitatively the powder deposition process.
It is implemented in the open-source package \acrshort{liggghts}~\cite{Kloss2012} (version 3.8.0), a \gls{md} variant suitable for granular materials.

The system modeled encompasses $N=$~76~000 deformable \gls{3d} \acrfull{ps} particles of diameter $d=$~\SI{2}{\mm}, surrounded by an \gls{al} container. 
Due to computational constraints, considering equivalent setup size, the particle size chosen in the simulation is significantly larger than in experiment ($d=\SI{80}{\micro\meter}$, see Sec.~\ref{sec:experiment}). 
In order to represent strongly cohesive powders, we include a simple model for attractive interactions among particles, and concentrate on a qualitative rather than quantitative comparison.

In the model,
each point particle $i$ is
represented by a sphere, and overlaps with a particle $j$ by a distance of $\delta_{ij}$.
The Hertz-Mindlin contact model~\cite{Hertz1882, Mindlin1949, Mindlin1953}
is used for the force calculations:
each particle pair interacts through a
non-linear spring-dashpot viscoelastic mechanical response.
The force $\vec{F}_{ij}$ resulting from a collision 
is expressed as a function of the overlap $\delta_{ij}$ and relative velocity through its normal $v_{n_{ij}}$ and tangential $v_{t_{ij}}$ components:
\begin{multline}\label{eq:HertzMindlin}
    \Vec{F}_{ij} =  \Big(   
        k_{n} \, {\delta_{n_{ij}}}^{3/2} -
        \gamma_{n} \, v_{n_{ij}} \, {\delta_{n_{ij}}}^{1/4}
    \Big) \, \Vec{n}_{ij}  + \\
    \Big(
        k_{t} \, \delta_{t_{ij}} \, {\delta_{n_{ij}}}^{1/2} - 
        \gamma_{t} \, v_{t_{ij}} \, {\delta_{n_{ij}}}^{1/4} \Big) \, \Vec{t}_{ij} -     
     \kappa \, A_{ij} \, \Vec{n}_{ij}.
\end{multline}
The first two terms of Eq.~\ref{eq:HertzMindlin} are the normal and tangential components of the force governed by the stiffness parameters
$k_{n,t}$ and viscoelastic damping parameters $\gamma_{n,t}$.
They represent the mechanical properties of the material constituting the particles.
While their numerical value can be linked to true material properties (elasticity modulus and Poisson ratio),  
they are bounded by numerical constraints.
Notably, optimization of simulation time requires to fix a sufficiently large time step $dt$;
but deeming $dt$ too large would result in overlooking certain collisions, thus invalidating the simulation. 
It is customary to consequently adapt the value of $k_{n,t}$ to remain at the lower end of the permissible spectrum, hence reducing computational effort while maintaining expected effects on the large scale. 
In the light of previous studies~\cite{Lommen2014, Vescovi2016, MALONE2008}, we estimate that
 $k_{n,t} \gtrsim 10^{5}$~N~m$^{-3/2}$ is sufficient to obtain stiff particle behavior using a time step $dt = 5 \cdot 10^{-7}$~\si{\s}. 
To reproduce inelastic particles, during a collision, most of the energy should be dissipated by viscous damping or through friction between the particles, which is obtained in the over-damped regime, once $\gamma_{n,t} \gg \sqrt{k_{n,t} \, m}$, which is by far the case with $\gamma_{n,t}  \sim 10^{8}$~kg~m$^{-1/2}$~s$^{-1}$.
The exact parameters used are given in Tab.~\ref{tab:sim_parameters}.
The tangential term is curbed to respect $|F_{t_{ij}}|  \leq \mu |F_{n_{ij}}| $, where $\mu$ is the 
friction coefficient (here $\mu=0.3$ for \acrshort{ps}-\acrshort{ps} contact) to account for frictional interactions between particles.
Rolling friction is also implemented, and provides an additional torque to the particles~\cite{Ai2011}.

\begin{table}
    \centering
    \scriptsize
    \renewcommand{\arraystretch}{1.2} 
    \begin{tabular}[c]{L{0.04\linewidth}L{0.52\linewidth}C{0.13\linewidth}C{0.12\linewidth}}
    \toprule
    {} & {Numerical parameters}  & {\textbf{\acrshort{ps}-\acrshort{ps}}} & {\textbf{\acrshort{ps}-\acrshort{al}}}  \\
    \midrule
    {$k_{n}$}  &  {Normal elastic coef.}	& {$1.52 \cdot 10^6$}        & {$4.10 \cdot 10^6$}    \\ 
    {$k_{t}$}    &  {Tangential elastic coef.}	& {$2.04 \cdot 10^5$}        & {$5.51 \cdot 10^5$}     \\ 
    {$\gamma_{n}$}  &  {Normal viscoelastic damping coef.}  & {$4.56 \cdot 10^{8}$}        & {$8.92 \cdot 10^{8}$}     \\ 
    {$\gamma_{t}$}  &   {Tangential viscoelastic damping coef.}	& {$4.10 \cdot 10^{8}$}        & {$8.01 \cdot 10^{8}$}    \\
    \bottomrule
    \end{tabular}
    \caption{Simulation parameters $k_{n,t}$ and $\gamma_{n,t}$ for the two types of interactions present in our model: among polystyrene particles (PS-PS) and between polystyrene particles and aluminum container (PS-Al). Elastic coefficients $k_{n, \, t}$ are in N~m$^{-3/2}$ and viscoelastic damping coefficients $\gamma_{n, \,t}$ in kg~m$^{-1/4}$~s$^{-1}$.}
    \label{tab:sim_parameters}	
\end{table}

The third term, $- \kappa \, A_{ij} \, \Vec{n}_{ij}$,
where $A_{ij}$ is the disc-shaped contact area between spherical particles $i$ and $j$, and $\kappa$ the cohesion energy density,
provides an extra cohesive component to the force calculation, in the form of a \gls{sjkr} model (based on the corresponding model of solid adhesion~\cite{SJKR1971}). 
It appends an additional attractive normal force contribution to the force calculation at each collision: as two particles enter into contact, this supplementary force tends to maintain the contact proportionally to the contact area, calculated from the overlap.
The cohesion energy density $\kappa$ represents all the cohesive forces between the particles, due to the reduction in surface free energy when particles are in contact. It encompasses many possible mechanisms responsible for cohesion in granular materials.
We hence use $\kappa$ as a proxy to tune the powder flowability in the simulation; values
from $\kappa = 10^{-4}$ to $\SI{e5}{\newton\per\meter\squared}$ are used to cover a wide range of interparticle-attraction strengths (see Fig.~\ref{fig:cohesion_simu}). 

The boundary conditions are embodied by contact surfaces following the stepwise powder deposition process described in Sec.~\ref{sec:process} (Fig.~\ref{fig:simplified_schematic_art}), with the dimensions of the apparatus used for experiment (see Fig.~\ref{fig:proto_cad}). 
The consecutive 
motions of each of those parts are listed in Tab.~\ref{tab:mesh_movements}.
The number of particles is calculated to fill the volume of the experimental container with a packing fraction $\varphi=0.6$, slightly lower than \gls{rcp} of frictionless monodispersed material~\cite{OHern2002}.
To achieve a realistic insertion of the particles inside the complex geometry, this step (Move~0 in Tab.~\ref{tab:mesh_movements}) is done by free fall of the particles inside the container (so-called sequential generation of \gls{rcp}), followed by the descend of the closing disc, which seals the container.

\begin{table*} 
    \centering
        \caption{Description of the consecutive motion of each element of the powder container modeled in the \gls{dem} simulation, corresponding to the \gls{3d} printing procedure described. Move 2.a and 2.b occur simultaneously. $d$ is one particle diameter. Time steps are given in simulation units, $dt= 5 \cdot 10^{-7}$~\si{\s}.}
    \scriptsize
    \renewcommand{\arraystretch}{1.2} 
    \begin{tabular}[c]{L{0.11\textwidth}L{0.07\textwidth}L{0.12\textwidth}L{0.42\textwidth}L{0.11\textwidth}}
    \toprule
    {Time step (end~of move)}  & {Move} & {Part} & {Movement} & {Velocity} \\
    \hline
    {$1.0 \cdot 10^{6}$}  & {Move 0} & {Closing disc} & {Linear movement in $-z$-direction: close printing bed} & Speed \SI{0.23}{\meter\per\second} \\ \hline 
    
    {$1.2\cdot10^6$} & & & Particle settling (realized under $+1g$) & \\
								& & & Adapt $g$-level 			& \\
    
    {$1.4 \cdot 10^{6}$}  & {Move 1} & {Inner cylinder} & {Linear movement in $+z$-direction: incremental rise by $10 d$} & Speed \SI{0.20}{\meter\per\second} \\ \hline 
    {$1.6 \cdot 10^{6}$}  & {Move 2.a} & {Inner cylinder} & {Clockwise rotation about $z$-axis} & {Period \SI{0.025}{\second}}  \\
    
    & {Move 2.b} & {Closing disc} & {Linear movement in $-z$-direction} & Speed \SI{0.05}{\meter\per\second} \\
    
    {$1.8 \cdot 10^{6}$}  & {Move 3} & {Closing disc} & {Counterclockwise rotation about $z$-axis} & {Period \SI{0.05}{\second}} \\
    
    {$2.0 \cdot 10^{6}$}  & {Move 4} & {Closing disc} & {Clockwise rotation about $z$-axis} & Period \SI{0.05}{\second} \\
    
    {$2.2 \cdot 10^{6}$}  & {Move 5} & {Outer tube } & {Clockwise rotation about $z$-axis} & {Period \SI{0.025}{\second}}  \\ \hline
    
    {$2.6 \cdot 10^{6}$}  & {Move 6} & {Vibrating disc} & {Shaking along $x$-direction} & {Period \SI{0.05}{\second}} \\
    
    {$3.0 \cdot 10^{6}$}  & {Move 7} & {Vibrating disc} & {Shaking along $y$-direction} & {Period \SI{0.05}{\second}} \\ \hline
    
    {$3.2 \cdot 10^{6}$}  & {Move 8} & {Inner cylinder} & {Linear movement in $-z$-direction: compression by $5d$} & {Speed \SI{0.1}{\meter\per\second}} \\ 
    \bottomrule
    \end{tabular}
    \label{tab:mesh_movements}
\end{table*}

Snapshots from the visualization of the simulation results are presented in Fig.~\ref{fig:snapshots}, and a video is available as video~1 of the supplementary material \cite{sm}.

\begin{figure}
	\centering
	\includegraphics[width=\linewidth]{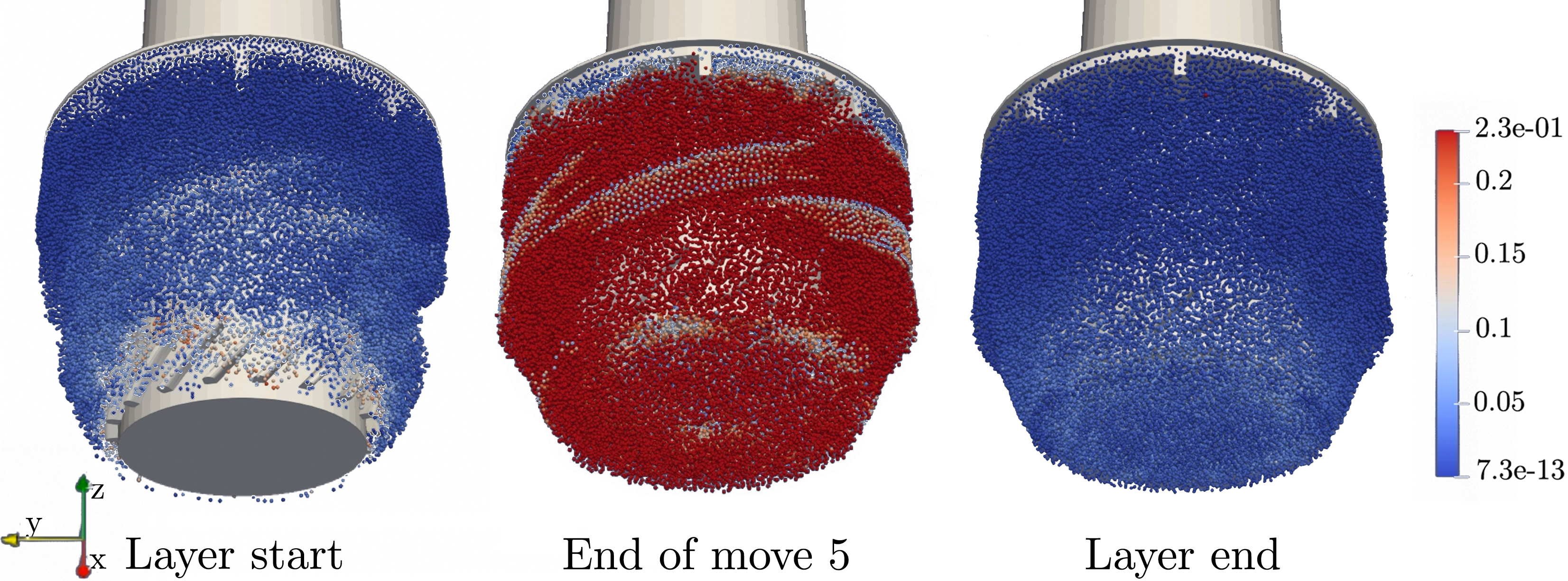}
	\caption{Snapshots of the simulation results: on the left, at the beginning of the process, before the inner cylinder rise; in the middle at the end of move~5, corresponding to the end of powder transport; on the right at the end of the deposition of one layer. The particles' color indicates their velocity magnitude, made explicit by the scale bar on the right (in \si{\meter\per\second}).}
	\label{fig:snapshots}
\end{figure}

\subsection{Procedures for data analysis}
 
The critical region regarding powder deposition quality
is the centre of the powder layer at the bottom of the processing container, which will be selectively solidified.
In simulation, this region is a disc 
of radius $15 d$ (where $d$ is the diameter of one particle),
and height $5d$ after the compression step. 
Analysis of the particles' distribution within this region is carried out by finding the local packing fraction $\varphi$ associated to each particle
from a Vorono\"{i} tessellation~\cite{Voronoi1908}:
each particle is assigned a unique flat-faced polyhedron representing the region of space closer to the centre-point of each particle than to the centre of any other particle. 
The ratio of particle volume and polyhedron volume
represents the local packing fraction~\cite{Schaller2015}. 
The mean local packing fraction $\left< \varphi \right>$ can hence be found for any region of space.
Vorono\"{i} tessellation is performed by the Python package SciPy~\cite{Virtanen2020}
and verified with the Voro++ open source software library~\cite{Rycroft2009};
ambiguous Vorono\"{i} cells (e.g.~at the system's boundaries) are discarded.
To visualize the simulation results, the bottom layer of the powder bed is divided into concentric rings, each equally spaced by $2d$. 
This division of the powder bed bottom layer is represented in Fig.~\ref{fig:plot_gravity}(b), the printing region represented in grey.

\subsection{Simulation results: influence of gravity}

\begin{figure}
	\includegraphics[width=\linewidth]{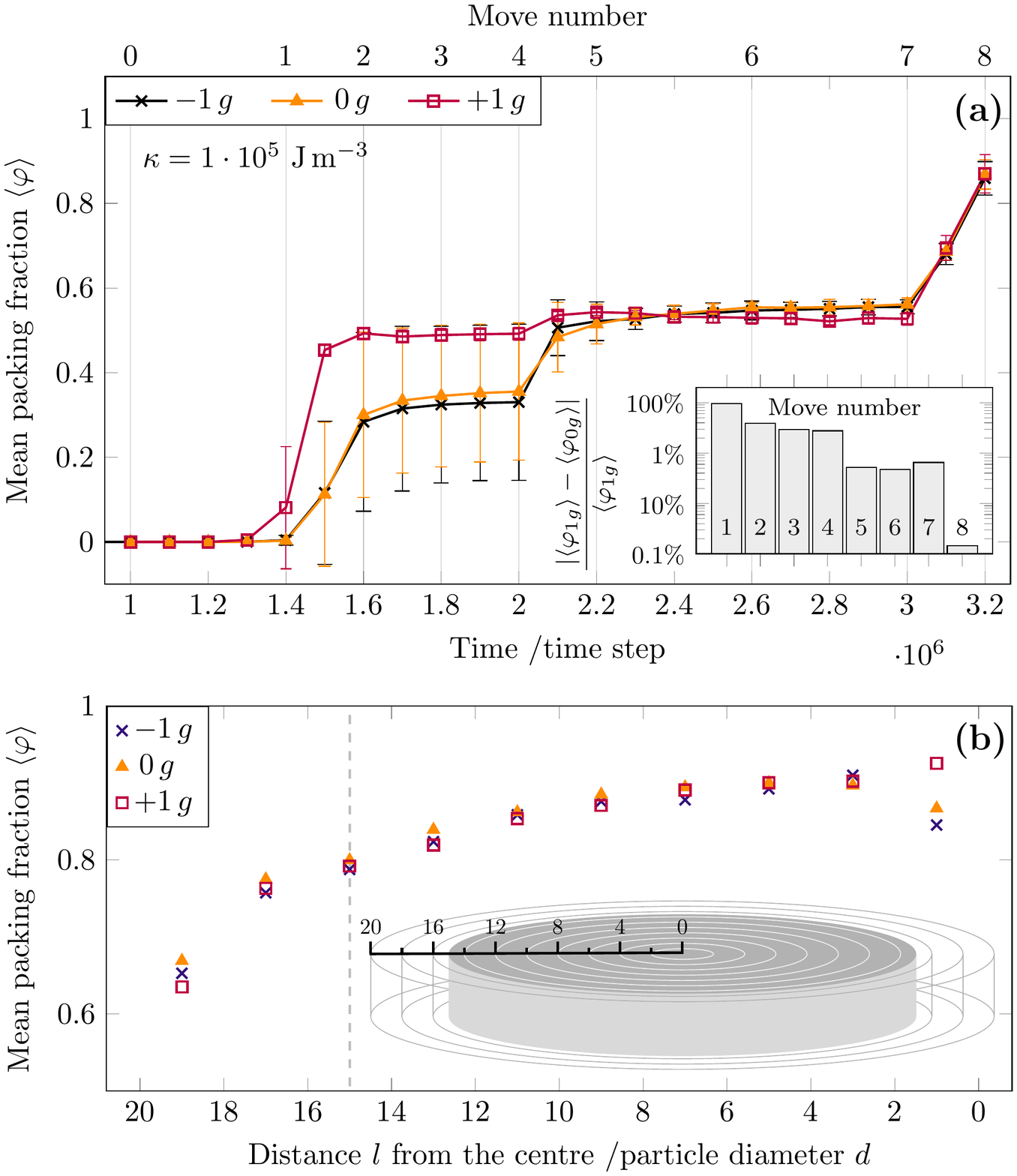}
	\caption{(a)~Mean packing fraction $\left< \varphi \right>$ on the printing region after each move of the powder deposition process (moves~0 to 8), for gravitational acceleration	
$+1g$, $0g$ and $-1g$ ($g=\SI{9.81}{\meter\per\second\squared}$). The local packing fraction $\varphi$ is obtained for each particle by Vorono\"{i} tessellation, error bars represent the standard deviation of the $\varphi$ distribution over the entire printing region. The inset shows the normalized difference between $+1g$ and $0g$ at the end of each move, expressed in percentage.
	 (b)~$\left< \varphi \right>$ per ring of the bottom layer at the end of move~8 (end of layer deposition), for $+1g$, $0g$ and $-1g$. The division of the powder bed bottom layer in concentric rings, equally spaced by two particle diameters $d$, is represented as an inset. The printing region is only the middle cylinder, represented in grey (and marked by a vertical dashed line).
	 }
	\label{fig:plot_gravity}
\end{figure}

The influence of the gravitational environment is studied by adjusting\ the gravitational constant $g=$~\SI{9.81}{\meter\per\second}, by multiplying it by $+1$, $0$ and $-1$ (labels used in the subsequent text and in Fig.~\ref{fig:plot_gravity} are respectively $+1g$, $0g$ and $-1g$).
Under $0g$, no mass-dependent external force field is applied: particle flow is induced solely
by boundary motion and forces transmitted through surrounding particles. 
The $+1g$ environment promotes the fall of particles toward the bottom of the process container, whereas the $-1g$ condition tends to pull the particles towards the container's top, working against the desired flow direction.

\begin{figure*}
	\centering
	\includegraphics[width=\linewidth]{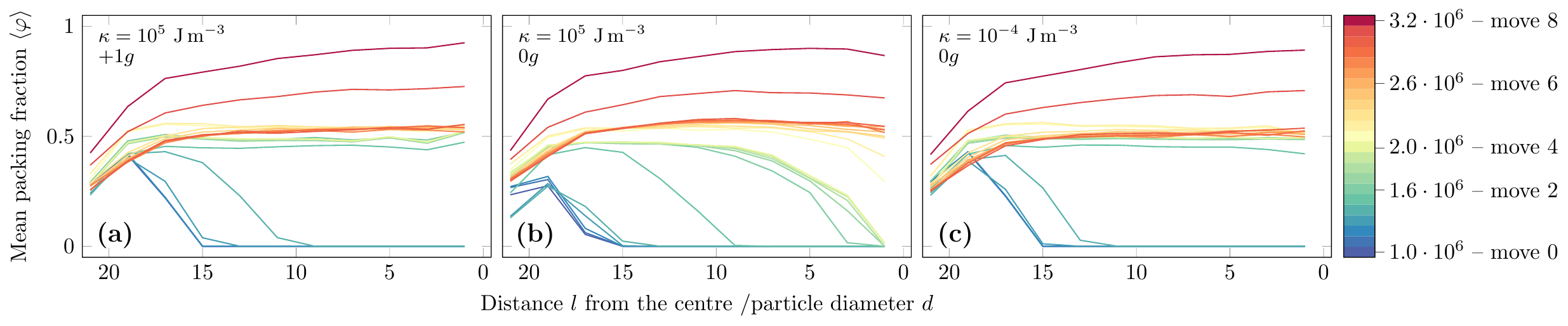}
	\caption{Mean packing fraction $\left< \varphi \right>$ per concentric ring (equally spaced by $2d$) as a function of the distance from the bottom layer centre (at abscissa 0) to its outer bound (at $20$), expressed in particle diameter $d$. The color code corresponds to the time $t$ given in simulation time steps.}
	\label{fig:distance_from_center}
\end{figure*}

The evolution of $\left< \varphi \right>$ at the end of each step of the powder deposition process, averaged over the printing region in Fig.~\ref{fig:plot_gravity}(a), shows that
the gravitational vector strongly modifies the granular density after the rise of the inner cylinder (move~1),
before the container's motion begins to shift material downwards:
under $+1g$, the particles' weight creates a collective motion towards the bottom, 
and then horizontally redistributes the particles as they slide on each other, piling until they reach a slope corresponding to the angle of repose.
The empty space under the printing platform is filled at $\left< \varphi \right> \approx 0.08$ solely under the effect of gravity at the boundary of the printing region, which can be observed in Fig.~\ref{fig:distance_from_center}(a): at $t=1.4\cdot10^{6}$ the outer part of the bottom layer, at a distance $l >15d$, is already filled at $\left< \varphi \right> \approx 0.41$, and particles have reached the outer diameter of the printing region, with $\left< \varphi \right>(10d \leq l \leq 15d) \approx 0.23 $, while this region is completely empty of particles under $0g$ -- see Fig.~\ref{fig:distance_from_center}(b).
In contrast, 
under $0g$ and $-1g$, the particles are not pushed towards this empty space, 
respectively due to a lack of mass-dependent force or a force towards the top of the container.
With move~2 begins the powder transport phase, which
triggers collective granular downward motion, regardless of the gravity-level. $\left< \varphi \right>$ undergoes a steep increase, but 
the normalized difference in packing fractions between $+1g$ and $0g$
-- represented in the inset of Fig.~\ref{fig:plot_gravity}(a) -- remains at 40\%. 
Throughout this transport phase (moves~2 to~5), 
material pushed downwards
slowly invades the printing region, already creating a relatively homogeneous layer under $+1g$, and remaining at the outskirts in $0g$, see Figs~\ref{fig:distance_from_center}(a) and~\ref{fig:distance_from_center}(b).
In Fig.~\ref{fig:plot_gravity}(a), the large standard deviation in $ \varphi $ results from the variability throughout the printing region:
as material is slowly pushed downwards, the centre remains at low or null packing fraction -- see Fig.~\ref{fig:distance_from_center}(b).
The normalized difference between $+1g$ and $0g$ has dropped to 30\% by move~4.
During move 5, the screw conveyor transports a large quantity of material downwards,
which is forced towards the container's bottom: some is pushed towards the centre, reducing the difference in $\left< \varphi \right>$ between $+1g$ and $0g$ to approx.~5\%.
The homogenization phase follows (moves~6 and~7):
the granular density becomes homogeneous throughout the printing substrate ($l \leq15d$ in Figs.~\ref{fig:distance_from_center}(a) and~\ref{fig:distance_from_center}(b)), regardless of the gravitational environment (see also Fig.~\ref{fig:plot_gravity}), culminating for all $g$-levels at $\left< \varphi \right> \approx 0.55$.
Finally, compression of the newly deposited layer (move~8) compacts the powder
and erases the remaining difference between $g$-levels, with $\frac{\left| \left< \varphi_{1g} \right> - \left< \varphi_{0g} \right> \right|}{ \left< \varphi_{1g} \right>} \approx$~0.15\%.
Zooming-in on the state of the bottom layer at the end of the powder deposition,
$\left< \varphi \right>$ after move~8 is shown in Fig.~\ref{fig:plot_gravity}(b) throughout the bottom layer.
The packing fraction within the printing region (for $l \leq 15d$) is very high, with $\left< \varphi \right> \approx 0.87$.
The average packing fraction achieved is significantly higher than the expected close packing of monodispersed spheres, indicating that particles overlap due to compression.
It is noteworthy that the most central region, being composed of the smallest volume, is more prone to statistical variability, which explains its stronger variation in $\left< \varphi \right>$ as a function of the $g$-level.
Besides, the closeness between the $0g$ and $-1g$ results is explained by the fact that we look solely at the bottom layer centre,
enforcing the importance of the horizontal motion (layer homogenization).

To summarize, the quality of the final powder deposition shows no dependence on the gravitational environment, although the simulations show, for the parameters chosen, that the powder transport does depend on the $g$-level. This suggests that the process is robust against changes in gravitational environment.

\subsection{Simulation results: influence of interparticle cohesion}

\begin{figure}
	\includegraphics[width=\linewidth]{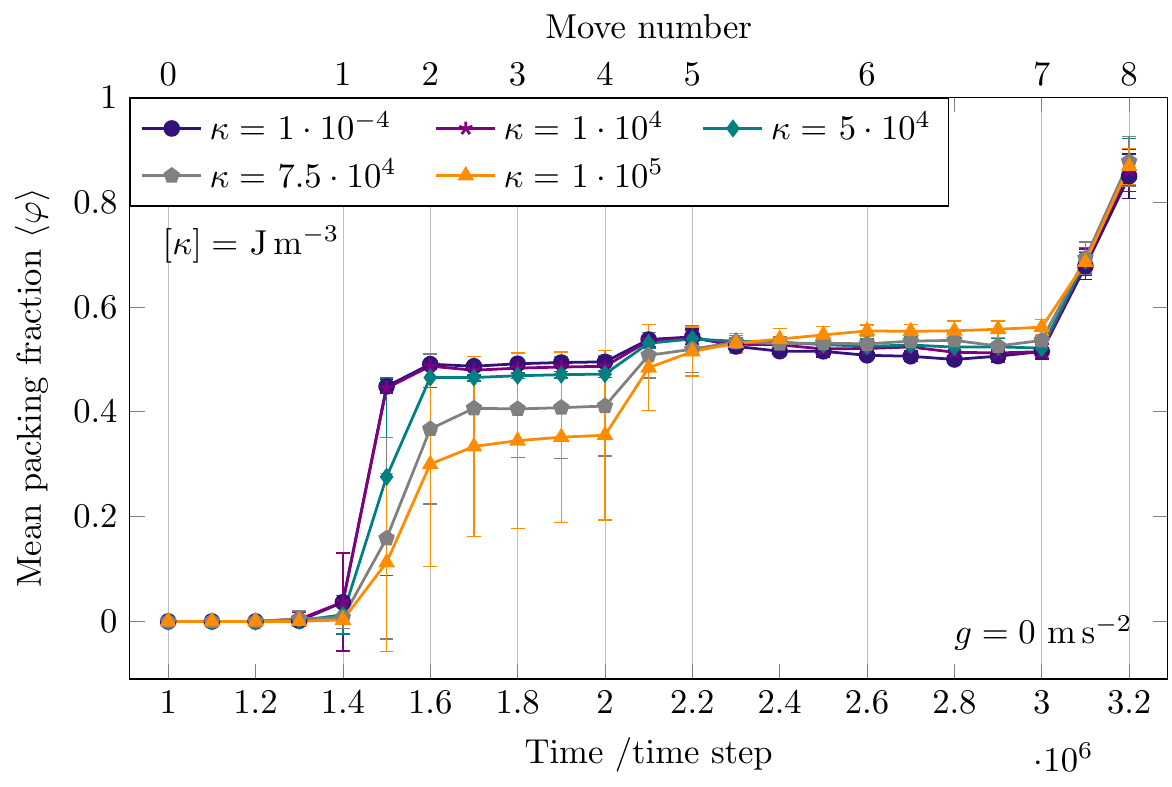}
	\caption{Mean packing fraction $\left< \varphi \right>$ on the printing region after each move of the powder deposition process (moves~0 to 8), for values of the cohesion parameter $\kappa$ ranging from $10^{-4}$ to $10^{5}$~\si{\joule\per\meter\cubed}, the extreme cases available in the present simulation.}
	\label{fig:cohesion_simu}
\end{figure}

The other relevant parameter for the \gls{am} process is dependence on the flow-properties of the powder feedstock. The corresponding simulation variable is the cohesion energy $\kappa$, which adds to the contact force an attractive term along the axis defined by the aligned particles centers, proportional to $\kappa$ times the area of contact $A_{ij}$
(see Eq.~\ref{eq:HertzMindlin}).
Numerical values of $\kappa$ are varied from $10^{-4}$~\si{\joule\per\meter\cubed} (for a very low cohesion, hence highly flowable powder) to $10^{5}$~\si{\joule\per\meter\cubed} (for a highly cohesive powder).
$\left< \varphi \right>$ throughout the powder deposition is presented in
Fig.~\ref{fig:cohesion_simu}. 

For all $\kappa \leq 10^{4}$~\si{\joule\per\meter\cubed}, our model shows very similar results:
such high flowability powder flows under the printing platform quickly, 
as it is already put into motion by the concurrent inner cylinder rotation and closing disc descend (move~2). Non-cohesive powders are very sensitive to collisions and tend to move freely within the container.
The remaining steps of the process have little effect on the powder repartition, 
with the exception of move~5, which slightly increases powder density in the printing region by bringing some more material downward, and the compression step (move~8) which significantly increases $\left< \varphi \right>$ from 0.52 to 0.85.
Locally measured $\left< \varphi \right>$ shown in Fig.~\ref{fig:distance_from_center}(c) reveal the same trend, as the transition between the initial state with the absence of powder at $l \leq 15d$ and the post-transport stage where $\left< \varphi \right> \approx 0.5$ at $l \leq 15d$ happens in less than $0.2 \cdot 10^{6}$ time steps.

The highest interparticular cohesion available in our model is $\kappa = 10^{5}$~\si{\joule\per\meter\cubed},
which in the simulation represents a low flowability powder.
Such highly cohesive particles tend to remain together, as outward forces resulting from collisions are minimized.
This is particularly visible on moves~2 to~5 in Fig.~\ref{fig:cohesion_simu}, where the difference between highest and lowest $\kappa$ reaches its maximum ($\approx 35\%$).
The efficiency of the powder downwards screw conveying (move~4) in microgravity is remarkable, closing the gap to bring all powders to $\left< \varphi \right> \approx 0.5$
In Fig.~\ref{fig:distance_from_center}(b),
the powder is pushed slowly downwards but does not invade the full printing region before being homogenized by shaking (moves~6--7).
This homogenization step brings the entire printing layer to $\left< \varphi \right> \approx 0.56$, slightly higher than for high flowability powders.
Again, the final layer compression 
completes the deposition process by fixing the particles in a state of high density, independently of their cohesive interactions: it erases 
all differences and brings the final $\left< \varphi \right>$ to $\approx 0.9$ for all materials.

To conclude this simulation study,
using the same \gls{3d} printing parameter set, 
all the simulation results have shown that 
whereas increased powder cohesion, as well as reversed gravity or absence thereof, modifies the powder flow behavior at each step of the powder transport taken separately, the general powder deposition functions independently of the raw material's flowability (cf.\ Fig.~\ref{fig:cohesion_simu}) and of the gravitational level (cf.\ Fig.~\ref{fig:plot_gravity}).
On average among all experiments discussed, it results in a final layer packed at $\left< \varphi \right> \approx 0.87$,
with a standard deviation among all experiments of $0.01$ only, 
erasing both internal and external variability factors. 
Universally fixed printing parameters can hence be used in our \gls{mug} experimental campaign,
enabling acute assessment of the effect of feedstock flowability decrease and $g$-level change.

%% file: Experiment.tex
Having confirmed by simulation the working principle of the 
proposed \gls{am} process, the latter is
implemented in two \gls{3d} printers and tested on-ground and in weightlessness.
\Acrfullpl{pfc}
supported by \gls{dlr} and \gls{esa}
are used to conduct \gls{mug} experiments:
the 34\textsuperscript{th} \acrshort{dlr} \gls{pfc} in September 2019, used for testing the hardware in \gls{mug}, and the 72\textsuperscript{nd} \acrshort{esa} \gls{pfc} in November 2019 (in the context of the \acrshort{esa} Education FlyYourThesis! GrainPower project), during which five samples were successfully manufactured fully in weightlessness, providing a proof of concept for the \gls{am} process. 
\Glspl{pfc} allow,
by flying an airplane describing parabolic trajectories,
a period of free fall of about \SI{22}{\s}, 
which
allows to perform experiments
in weightlessness.
This \gls{mug} period is surrounded by hypergravity phases as the plane rises and swoops.
This maneuver is typically repeated thirty-one times per flight day, a campaign consisting of three to four flight days. 
The final experimental rack used to produce the weightlessness samples contains two \gls{3d} printers, used to each produce one sample per flight-day.

\subsection{Experimental set-up}

\begin{figure}
	\centering
	\includegraphics[width=\linewidth]{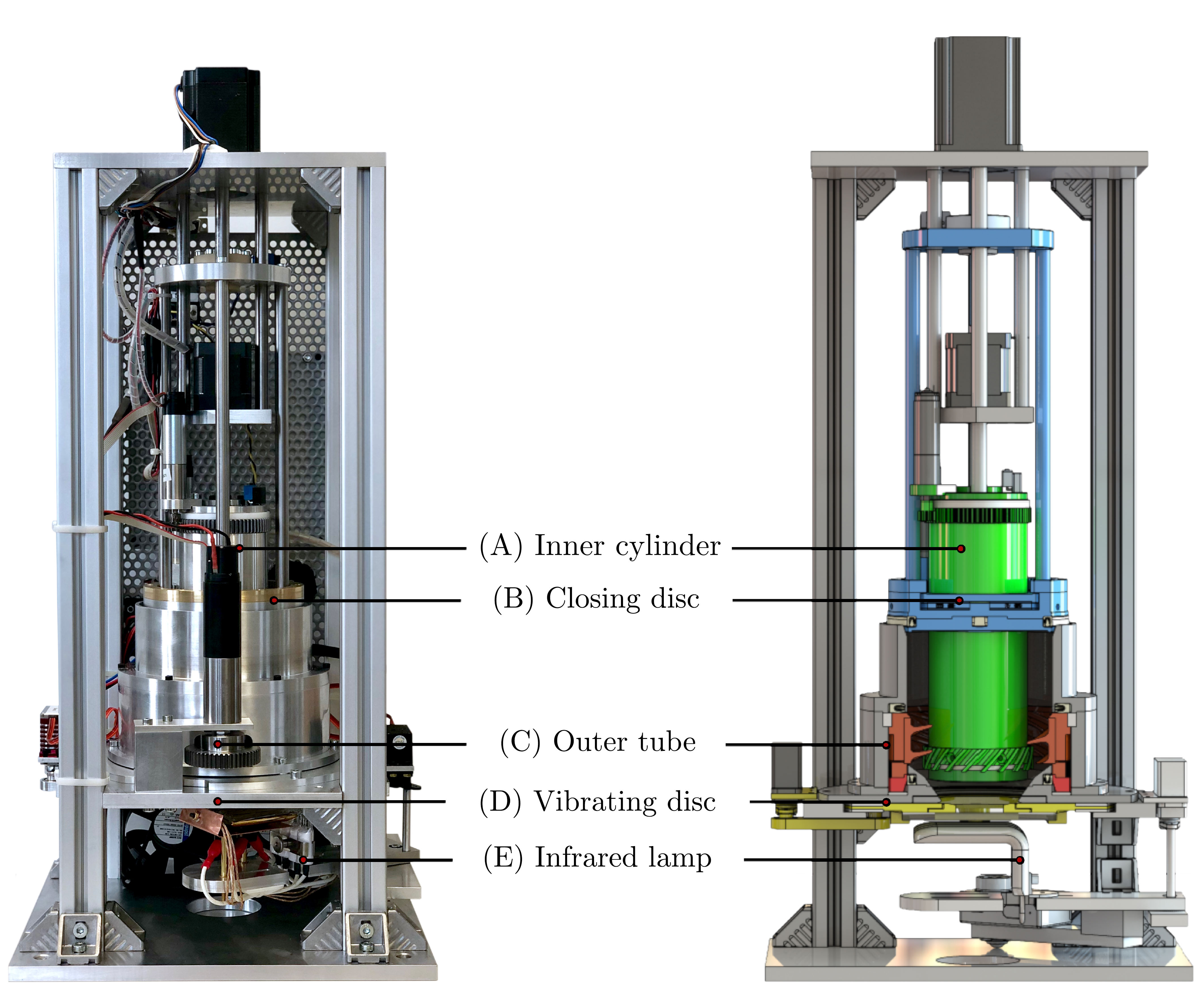}
	\caption{(Left) photography and (right) \acrfull{cad} of the \gls{3d} printer. The main components of the printer are labeled. On the \acrshort{cad}, the fixed structure is represented in grey while the moving parts are colored, each color representing a movement block.}
	\label{fig:proto_cad}
\end{figure}

A \gls{3d} printer built to implement the \gls{am} process described (and its digital counterpart) are shown in Fig.~\ref{fig:proto_cad}.
The powder container is composed of two coaxial cylinders: (A)~the inner cylinder and (C)~the outer tube, respectively of diameter~\SI{65}{\mm} and \SI{120}{\mm}.
From above, the container is enclosed by (B)~the closing disc, filling the space between the two cylinders and moving down to control the powder bed volume.
On the bottom, it is closed by (D)~the vibrating disc, which contains the solidification window, through which the raw material will be sintered by (E)~the energy source, placed underneath the powder container.
A thorough technical description is available elsewhere~\cite{Lopez2021thesis}. 

The type of energy source and the solidification window's material determine the maximum temperature allowed, hence the adequate raw materials. In the demonstration experiment presented here, polymer powders are used; an \gls{ir} lamp serves as energy source (Quattro \gls{ir} emitter from Heraeus, Germany), 
and the solidification window is a doubled \SI{5}{\mm} thick fused silica plate ({proQuarz}, Germany).
The general structure is formed by $30 \times 30$~\si{\milli\meter\squared} aluminum profiles on which the individual parts are mounted. 
The printing volume available is a cylinder of diameter \SI{65}{\mm} and height \SI{50}{\mm}.
The motion of each part exactly follows the description given in Fig.~\ref{fig:simplified_schematic_art} -- specific motion parameters are given in Tab.~\ref{tab:proto_parameters}.

\begin{table}
	\centering
	\caption{\label{tab:proto_parameters}\Gls{3d} printing parameters, $l$ being the layer height.}
	\scriptsize
    \renewcommand{\arraystretch}{1.4} 
    \begin{tabular}[c]{L{0.25\linewidth}L{0.16\linewidth}L{0.41\linewidth}}
    	\toprule	
    	\textbf{Part of printer} & \textbf{Motion} & {\textbf{Operating parameters}} \\
    	\midrule

    	\multirow{2}{*}{(A) Inner cylinder} 	& {Translation} 	& {$z$ steps: $+$\SI{1000}{\micro\m}, $-$\SI{500}{\micro\m}} ($l=\SI{500}{\micro\meter}$, compression ratio $\nicefrac{1}{2}$)\\
        													& Rotation		& Rotation speed \SI{0.102}{\m\per\s} \\

    	\multirow{2}{*}{(B) Closing disc} 	& Translation	& {$z$ step: $-$\SI{200}{\micro\m}}\\ 
    														& Rotation	 	& Rotation speed \SI{0.126}{\m\per\s}  \\ 
    	
    	(C) Outer tube 							& Rotation 		& Rotation speed \SI{0.226}{\m\per\s} \\
    	 
    	\multirow{2}{*}{(D) Vibrating disc} & \multirow{2}{*}{Shaking}		& Amplitude~\SI{2}{\mm} \\ 
    															& 	 											& Frequency~\SI{5}{\hertz} \\   
    	 \multirow{2}{*}{(E) Infrared lamp}	&\multirow{2}{*}{} 	& Lamp power \SI{500}{\watt}\\ 
    	 														&								& Sintering time \SI{20}{\s}\\
    	\bottomrule
    \end{tabular}
\end{table}

The \gls{3d} printing procedure is as follows:
first, the inner cylinder rises vertically to make space for the new layer. At the end of the powder deposition phase, it moves vertically downward to compress the powder underneath. To maintain a constant volume inside the powder containment, the closing disc translates down to counteract the motion of the inner cylinder; simultaneously, it rotates in alternating direction to probe the packing and keep it from jamming.
Powder downward transport is carried out by the 
screw conveyor placed inside the outer tube, rotating to push the powder towards the bottom of the container. The shell of the inner cylinder (excluding the printing substrate) also rotates (independently of its translation motion) to probe the powder flow from within the container, providing a rheological characterization of the feedstock powder.

As mentioned previously,
in this first study the
\textit{in-situ} probing control loop is not automatized
to allow better overview of the effects of our variable parameters (gravity and powder flowability) on the powder deposition.

\subsection{Material}

\begin{figure}[!h]
    \centering
        \subfigure[~\acrshort{rs} particle \label{subfig:roughfull}]{
    \includegraphics[width=0.48\linewidth]{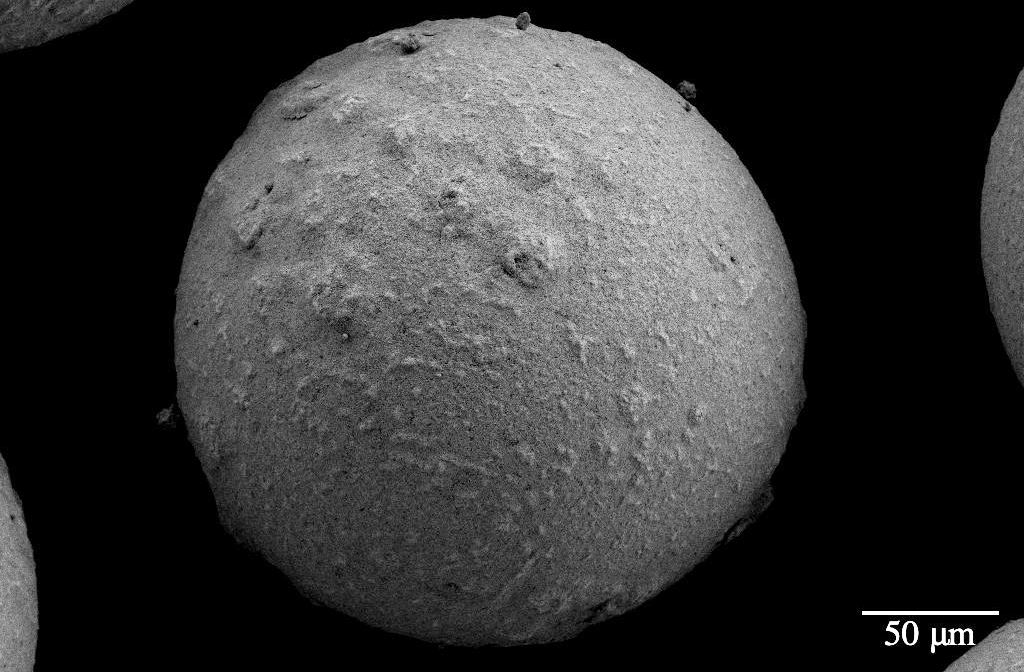}} \hfill
    \subfigure[~\acrshort{rs} particle closeup \label{subfig:roughsurf}]{
    \includegraphics[width=0.48\linewidth]{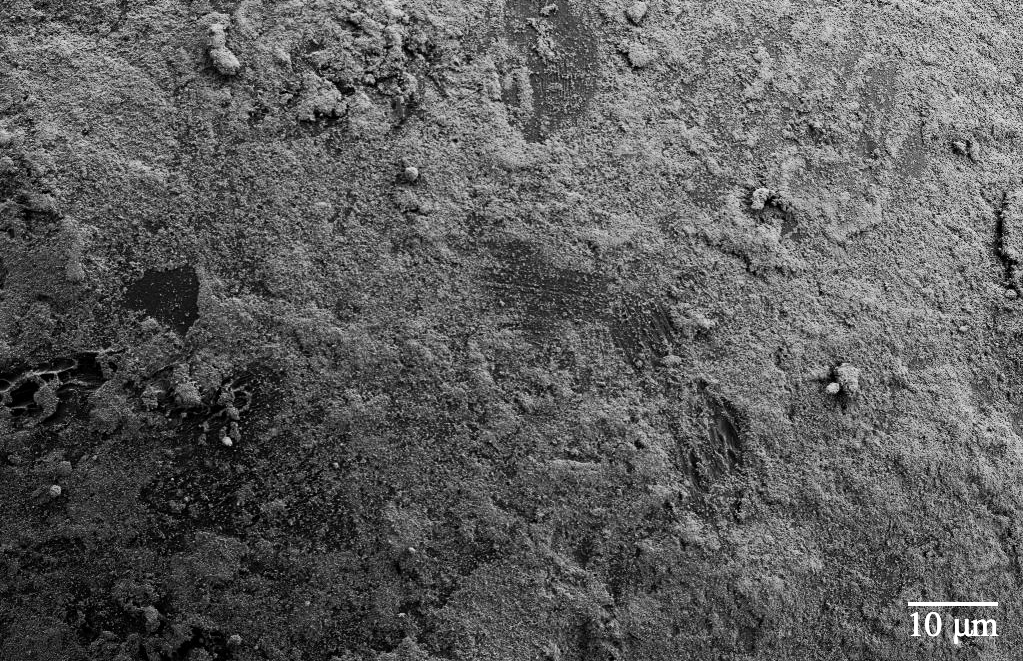}}\\
    \subfigure[~\acrshort{ss} particle \label{subfig:smoothfull}]{
    \includegraphics[width=0.48\linewidth]{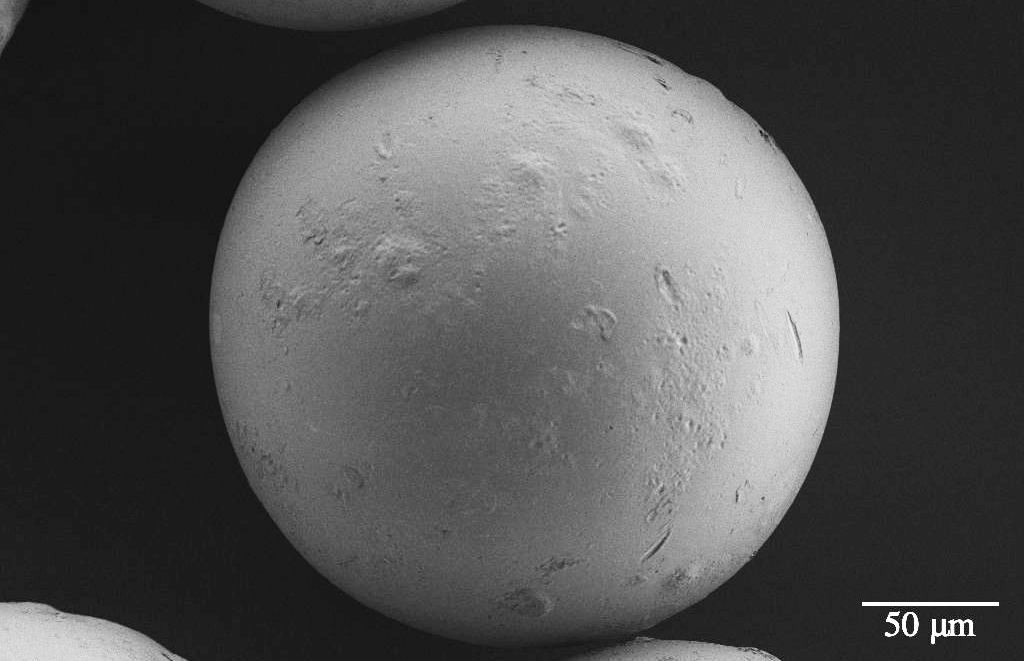}} \hfill
    \subfigure[~\acrshort{ss} particle closeup \label{subfig:smoothsurf}]{
    \includegraphics[width=0.48\linewidth]{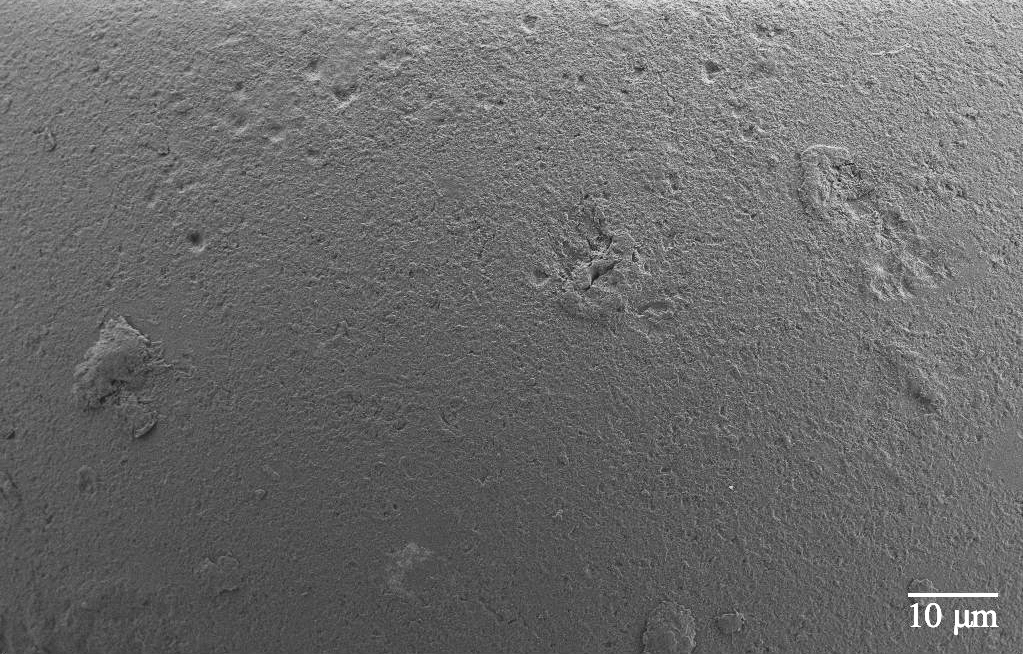}} 
	\caption{\Acrfull{sem} images of the \acrfull{ps} powder used as raw material (the microscopies are done for particles of diameter \SI{250}{\micro\meter}), (a) for powder with rough surface state (\acrshort{rs}), at the scale of a particle and (b) at the scale of its surface; (c) for powder with smooth surface state (\acrshort{ss}), at the scale of a particle and (d) at the scale of its surface. The increased surface roughness in the sub-micrometer range is clearly visible in (a) and (b), while (c) and (d) exhibit a much  smoother surface. \Gls{sem} imaging is done at \SI{1}{\kilo\electronvolt}.}
	\label{fig:SEM_surfacestates}
 \end{figure}

To demonstrate feasibility of \gls{am} from raw materials of high to mediocre flowability,
two exemplary demonstrator powders are used, which share all physical characteristics except for their surface roughness. This enables us to test solely the effect of a decreased flowability on the powder deposition and sintering process.
The model substances are crafted as follows. 
A monodisperse spherical \acrfull{ps} powder of main diameter \SI{80}{\micro\meter} is used 
(where size and dispersity are given by the manufacturer).
Produced by the company Microbeads under the name Dynoseeds,
the powder as-received from the manufacturer is dry-coated with a sub-micron angular \gls{ps} dust, as shown in Figs.~\ref{fig:SEM_surfacestates}a and \ref{fig:SEM_surfacestates}b.
This powder is labelled \gls{rs} in the subsequent text.
The surface coating is removed by wet-sieving the powder batch and subjecting it to ultrasound at a frequency of \SI{20}{\kilo\hertz} for a duration of 8 hours per batch. The resulting particle surface state is shown in Figs.~\ref{fig:SEM_surfacestates}c and \ref{fig:SEM_surfacestates}d: the asperities have been removed, leaving exposed the smooth surface of the spherical particles. This powder is named \gls{ss} in the subsequent text.

The influence of the surface coating on the flowability of the powder is not \emph{a priori} evident, as the addition of smaller guest particles on large host particles 
can either enhance or hinder their flowability.
A small quantity of hard
guest particles, of diameter much smaller that the host particles~\cite{Yang2005}, can enhance flow by reducing the curvature radius at the contact point, hence the Van-der-Waals attractive interactions~\cite{Castellanos2005} . However, it was shown empirically that
if the \gls{sac} of the host particles is greater than approximately 20\%, the inverse effect can be observed, \textit{viz.} a reduction of the flow properties as compared to the pure host particle~\cite{Castellanos2005, Fulchini2017}.
We thus expect our \gls{rs} particles to constitute a powder with reduced flowability compared to the \gls{ss} particles; in the following, we will also confirm this experimentally by flow energy tests.

\begin{figure}
    \centering
    \subfigure[Before testing\label{subfig:fresh_closeup}]{
    \includegraphics[width=0.48\linewidth]{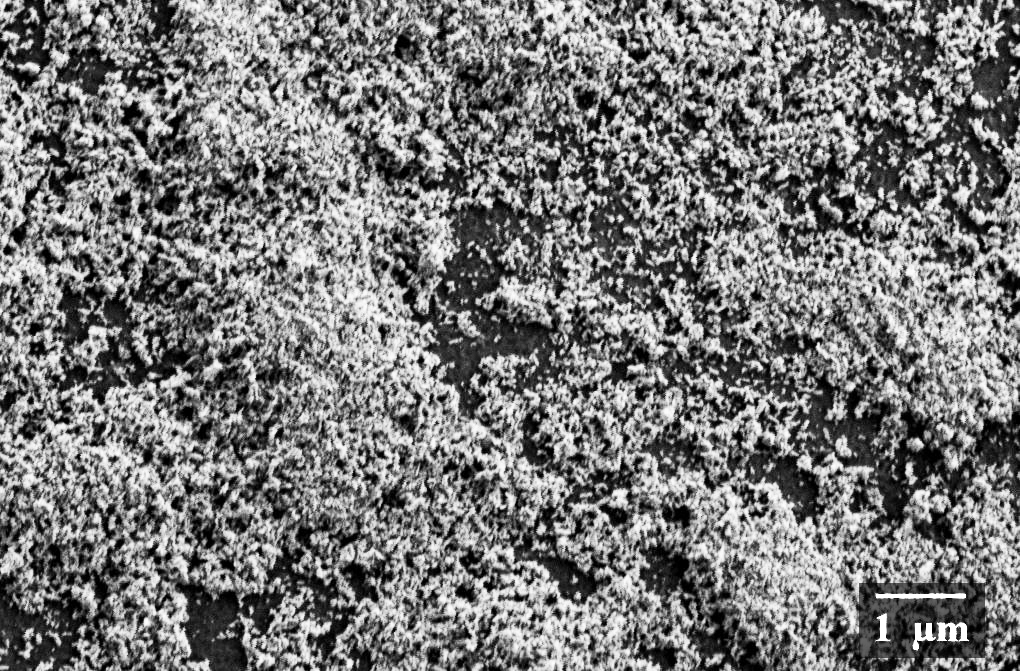}} \hfill
    \subfigure[After testing\label{subfig:fresh_used_closeup}]{
    \includegraphics[width=0.48\linewidth]{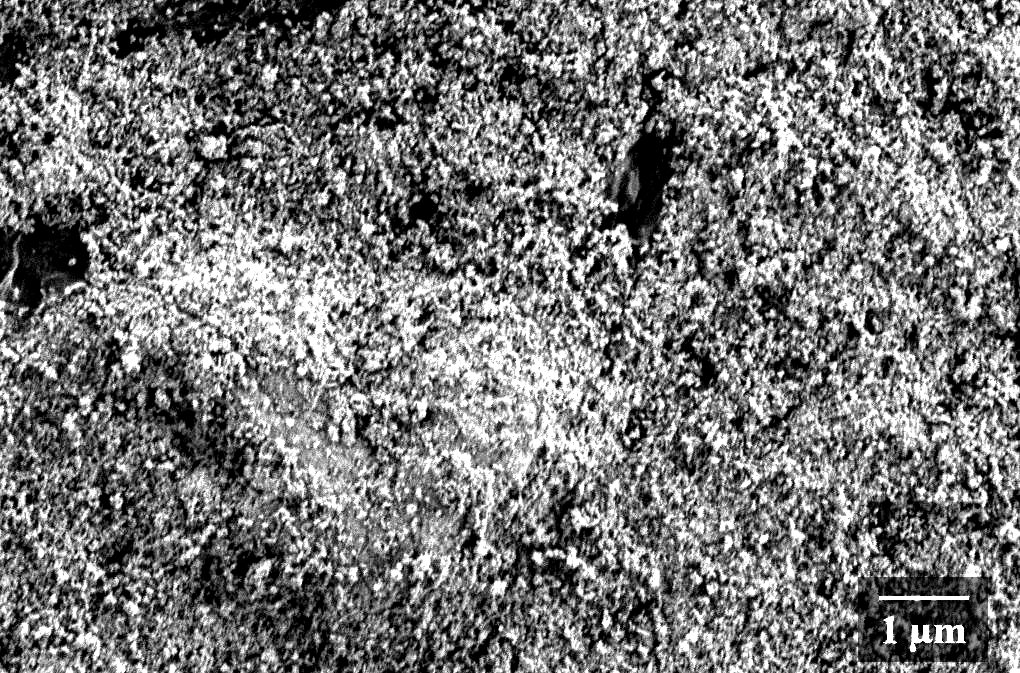}}
	\caption{\acrshort{sem} images of the \acrfull{rs} \acrlong{ps} powder used as raw material (a)~before and (b)~after a twenty hours long shear test (see text for details). The sub-micrometer surface roughness remains present in the same quantity and homogeneity on both images. Imaging is done at \SI{1}{\kilo\electronvolt}.}
	\label{fig:SEM_surfacestates_beforeafter}
 \end{figure}

The resistance to wear of the rough coating is tested to ensure its persistence throughout the experiment. 
To validate the principle of our experimental study, it is essential
that the model powders 
retain their flow-properties throughout the entire experiment, independently of the load to which they will be submitted during powder deposition.
To test this, we have subjected the
particles to shear in a Couette-Taylor shear cell \cite{Larson}, continuously for 20 hours at increasing shear rate, $\dot{\gamma}= 10^{-2}$ to $10^{3}$~\si{\per\s},
including air-fluidization before and after at a flow rate of \SI{5}{\liter\per\minute}; 
\gls{sem} microscopies taken before and after are shown in Fig.~\ref{fig:SEM_surfacestates_beforeafter}.
The surface is visibly unchanged by the long duration shear test: the rough coating is still distributed on the entire particle surface,
showing that it remains despite frictional contacts.

How to best characterize the rheology of powders for \gls{am} is an open question~\cite{Vock2019}.
The term \emph{powder flowability} is widely used and intuitively understood; however, it lacks a clear definition, 
as it does not rely on a normalized measurement method,
nor on international system units. 
A powder is defined as \emph{flowable} 
if it tends to plastically deform (i.e.~flow akin to a liquid) under a certain stimulus -- which
may simply be its own weight. 
In contrast, a \emph{non-flowable} powder resists flowing and tends to maintain its shape akin to a solid. If it is forced into flowing by an external load,
it will do so in large chunks of material themselves preserving their shape, in an erratic manner and showing higher tendency to block the flow by forming stable aggregates that can withstand a finite amount of stress before yielding (jammed regions). In other words, contacts between particles tend to be more enduring~\cite{Castellanos2005}.
The terms 
\emph{high} and \emph{low flowability} are used throughout the present work following this phenomenological definition.

\begin{figure}
\centering
\includegraphics[width=\linewidth]{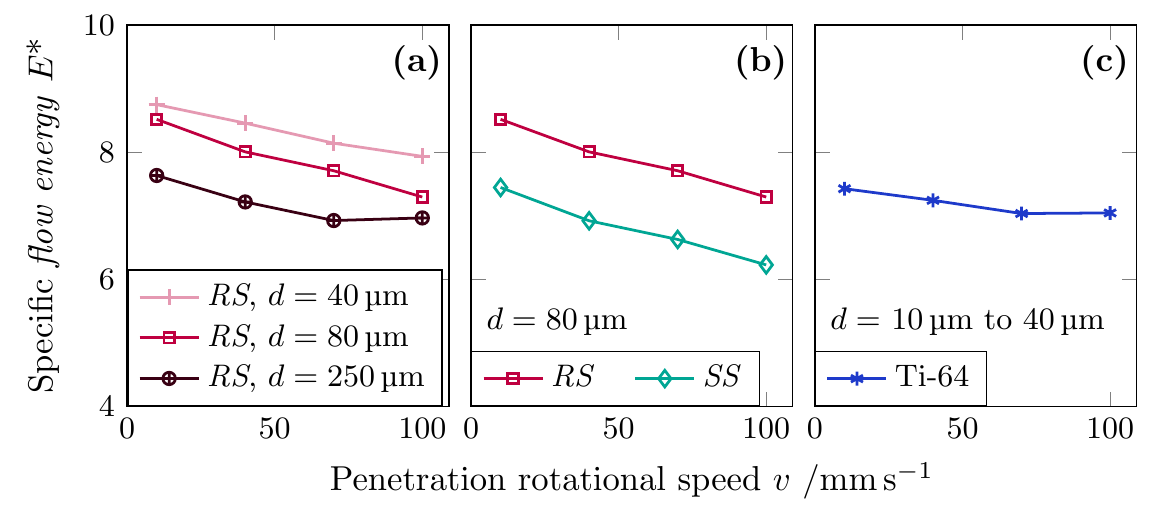}
\caption{Specific \textit{flow energy} $E\text{*}$ as a function of the helix speed $v$ for samples (a)~of \acrfull{rs} \acrlong{ps} powder of diameter \SI{40}{\micro\meter}, \SI{80}{\micro\meter} and \SI{250}{\micro\meter};
(b)~of \SI{80}{\micro\meter}-diameter \acrfull{ss} and \gls{rs} powders (model materials used for additive manufacturing experiment);
(c)~of polydisperse Ti-64 powder with size in the range \SI{10}{\micro\meter} to \SI{40}{\micro\meter}, a typical material for metallic powder-based \gls{am}~\cite{Liu2019}.
}\label{fig:flowenergy}
\end{figure}

In absence of a universal definition, 
the procedure that we used to characterize flowability of the \gls{ss} and \gls{rs} powders
is the so-called \textit{flow energy} measurement available on the \gls{ft4}~\cite{Leturia2014, Yang2015}. 
It consists of extracting from a powder bed of height $h$ an helix of angle $\alpha$ and radius $r$, at different speeds $v$, recording the torque $M$ and normal force $F$ to calculate the \textit{flow energy} $E$.
Following Wenguang et al.~\cite{Wenguang2017conf, Wenguang2017}, 
a dimensionless \textit{flow energy} $E\text{*}$
is introduced by normalizing
the \textit{flow energy} by the potential energy of the sample:
\begin{equation}
    E\text{*}=\frac{E}{m_s \, h \, g} = \frac{1}{m_s \, h \, g} \int_{0}^h \left(\frac{M(h')}{r \tan \alpha} + F(h') \right) dh',
    \label{eq:flowenergy}
\end{equation}
where $m_s$ is the total mass of the sample and $g$ the gravitational acceleration.
Qualitatively, an increase in \textit{flow energy} $E^*$ corresponds to a decrease in powder \textit{flowability}.
Fig.~\ref{fig:flowenergy} shows 
the specific \textit{flow energy} $E\text{*}$,
measured at penetration rotation speeds between 10 and~\SI{100}{\mm\per\s}
for powders at high packing fraction $\varphi =0.6$.
Different rotation speeds are used to ensure that the results are
robust; note that a speed of \SI{100}{\mm\per\s} is comparable to the
rotation speed used in our printers.
In an effort to contextualize our demonstrator powders,
the reader is provided with
material for comparison:
$E\text{*}$ is measured for 
\gls{rs} powders of smaller and larger particle diameter (respectively $d=$~\SI{40}{\micro\meter} and  $d=$~\SI{250}{\micro\meter}),
and for a Ti-64 metal alloy powder, typically used in \gls{am}.
The latter metal powder is used to compare the flow properties of the two polymer demonstrator powders to that of 
a commercial \gls{am} material, although the \gls{am} process presented
is not in its current form adapted to \gls{3d} print such material.

It has been shown 
that particles of smaller diameter tend to exhibit higher cohesion, due to the predominance of van der Waals interactions~\cite{Castellanos2005,Zimmermann2004,Israelachvili2011b,Heitmeier2021}.
Hence the corresponding powders undergo a flowability decrease.
This decrease is indeed captured by the specific \textit{flow energy} test shown 
in Fig.~\ref{fig:flowenergy}(a):
as the diameter $d$ doubles (40 to \SI{80}{\micro\meter}),
the \textit{flow energy} needed to make the powder flow decreases (by 5.7\% on average),
showing a \emph{better} flowability of the powder comprised of larger particles.
The particle diameter is then increased to \SI{250}{\micro\meter},
inducing a 
further decrease of $E\text{*}$ (by 9.6\% in average),
again showing that larger particles amount to a powder of \emph{higher} flowability.
The specific \textit{flow energy} increases with decreasing particle size; this trend is preserved over all the rotation speeds measured.

In Fig.~\ref{fig:flowenergy}(b),
$E\text{*}$ is presented for the \gls{ss} and \gls{rs} \SI{80}{\micro\meter} diameter powders
used in our experiment. 
In the dense packing of rough particles,
surface friction is activated as asperities on the particles' surfaces interlock:
the \textit{flow energy} is
15\% higher for the \gls{rs} powder.
The effect of surface roughness is clearly visible: increased friction begets higher stress necessary for particles to slide along each other, thus higher stress to trigger flow;
the \gls{rs} powder exhibits lower flowability than the \gls{ss} powder.
This effect is at least comparable to the one induced by a change of the particles' diameter by more than a factor 6.

Finally, a comparison is provided in Fig.~\ref{fig:flowenergy}(c) to a commercial \gls{3d} printing metal powder:
the polydisperse Ti-64 alloy powder~\cite{Liu2019}, with 
particle diameter in the range \SIrange{10}{40}{\micro\meter}.
Its flowability is slightly higher than the \gls{rs} powder but much lower than the \gls{ss} powder, placing our two demonstrator powders as boundaries framing 
a typical material used in powder-based \gls{am}, in terms of flow-behavior.
Note that the \gls{rs} powder of similarly small particle size would
be significantly less flowable than the Ti-64 powder at all rotation speeds.

\subsection{Manufacturing procedure \& sample characterization}

\paragraph{\Gls{3d} printing procedure}
For the sake of comparison, 
all samples are obtained using the same printing parameters.
The material deposition lasts \SI{20}{\s}.
The layer height is \SI{500}{\micro\meter} (corresponding to $\sim6d$) and the compression rate is 50\% --~i.e.~the printing platform rises by \SI{1000}{\micro\meter} before each deposition step, then descends by \SI{500}{\micro\meter} to compress the newly deposited layer.
Following the compression stage, which immobilizes the newly deposited layer, 
sintering takes place.
For the \gls{mug} samples, sintering starts at the next
parabola, to allow the entire manufacturing procedure to be carried out in weightlessness.
To provide as much information as possible on the powder deposition,
no further compression is applied during solidification, and sintering is preferred over melting, as it maintains possible heterogeneities of the deposited powder layer.
The sintering parameters were chosen such that at each new layer, a depth of \SI{1000}{\micro\meter} be sintered (i.e.~twice the layer height), to ensure full cohesion between the subsequent layers.

\begin{figure}
    \centering
    \subfigure[\label{subfig:sample_heatzone} Sample preparation rationale]{
	\includegraphics[width=0.93\linewidth]{sample_heatzones.pdf}} \\
	\subfigure[\label{subfig:IRlamp} \Gls{ir}-lamp high heat areas]{
	\includegraphics[height=0.32\linewidth,trim={60pt 120pt 60pt 30pt},clip]{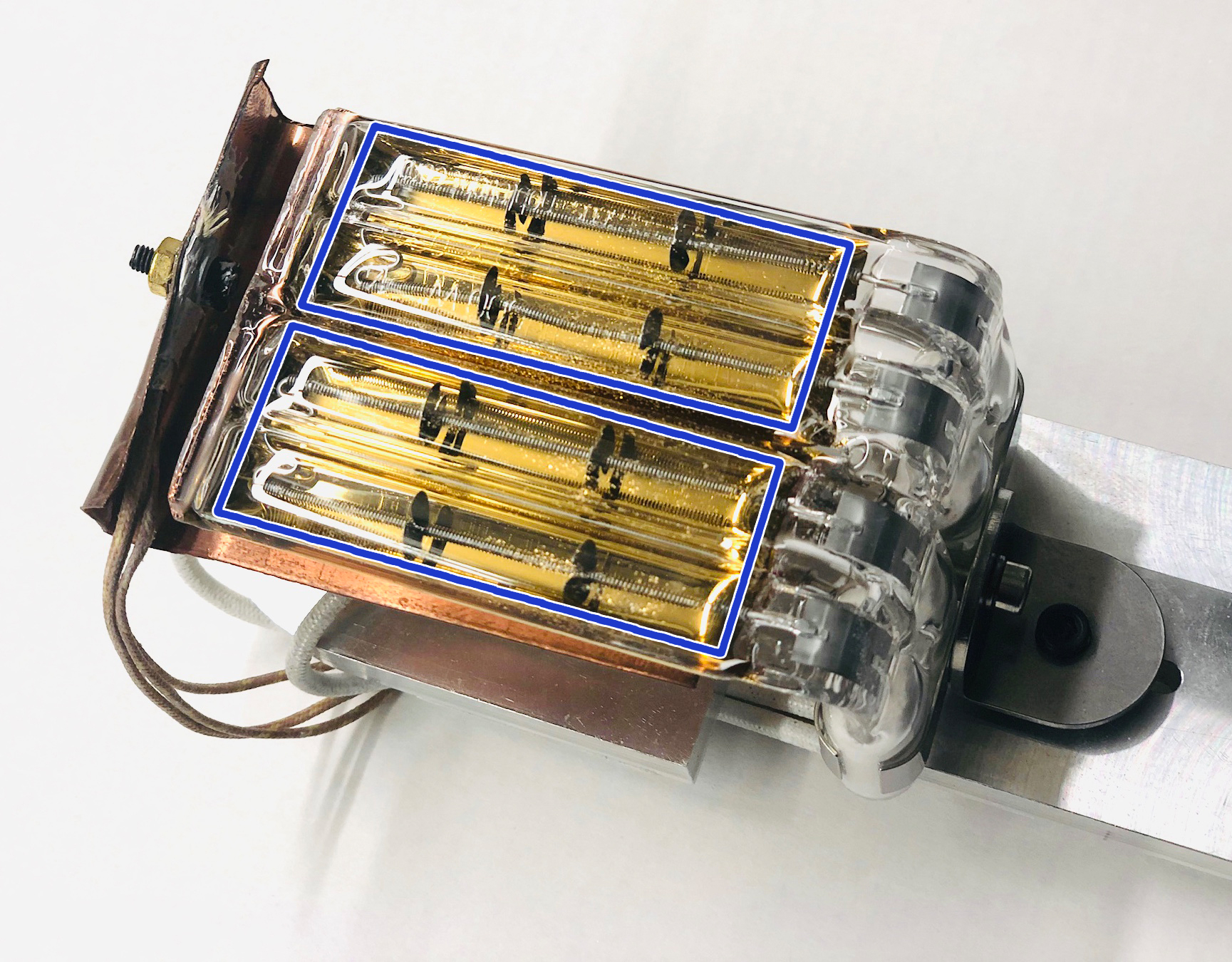}}
	\subfigure[\label{subfig:selecsample} Selected sample]{
	\includegraphics[height=0.32\linewidth,trim={0pt 10pt 10pt 20pt},clip]{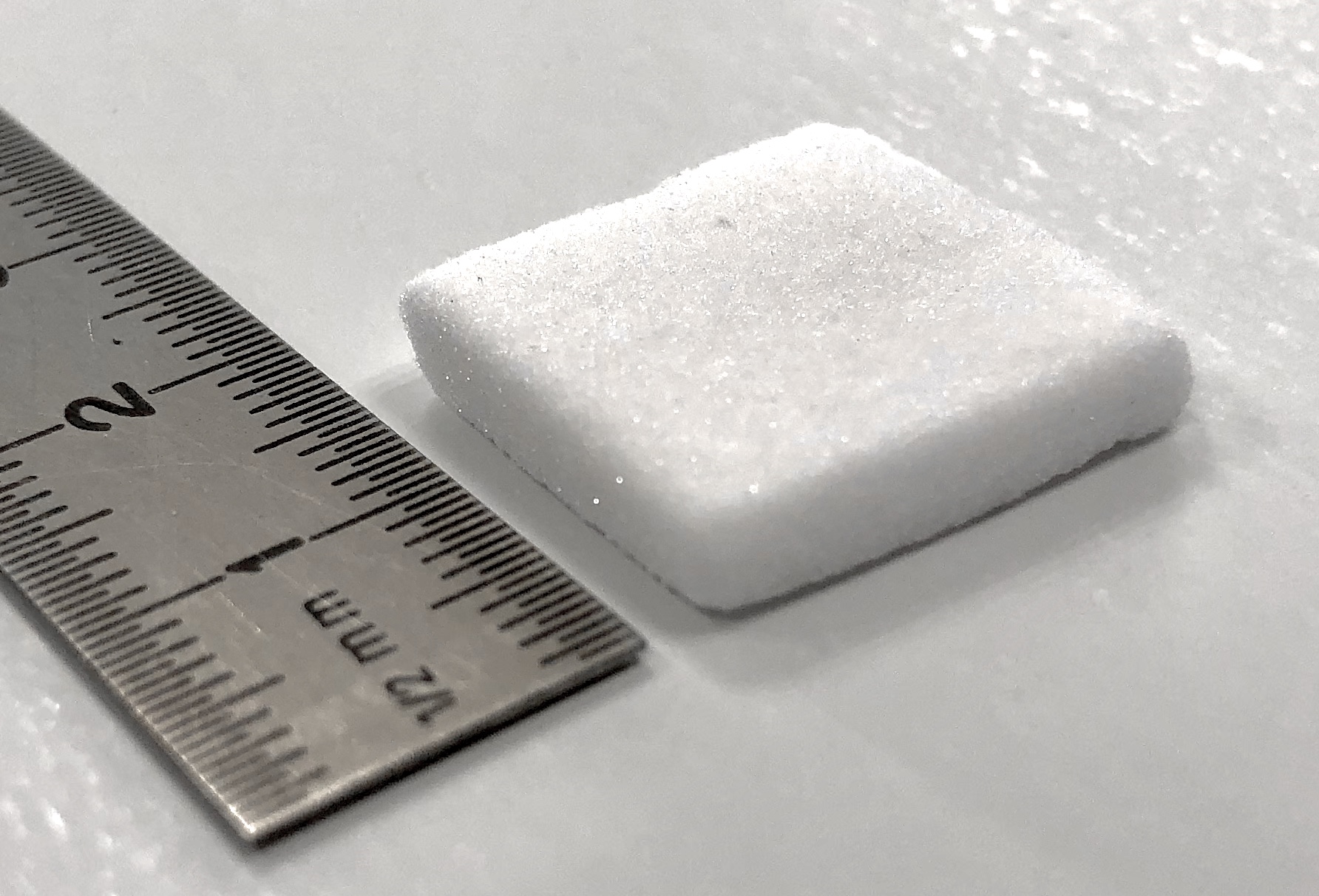}}
    \caption{Sample preparation rationale. (a)~Superposition of high heat zones with the centre of the printing area, including delimitation of the selected sample. The sample represented here is manufactured on-ground from \acrfull{rs} powder. (b)~\Gls{ir}-lamp including schematic representation of high heat zones (in blue on the picture), corresponding to the zones marked in blue on the full sample. (c)~Resulting sample after cutting (the sample represented here is sample $L3$, \gls{3d} printed in weightlessness from \acrfull{rs} powder). } 
    \label{fig:samples_photo}
\end{figure}

\paragraph{Printing substrate}
All samples were \gls{3d} printed on a pre-manufactured printing substrate. Examples of printing substrates are visible in Fig.~\ref{subfig:sample_heatzone} (circled in yellow and labeled \enquote{full printing area}) and Fig.~\ref{fig:samples}.
Substrates were manufactured by oven-sintering the same \gls{ps} powder as used in the experiments on an aluminum holder for approx.~\SI{1}{\hour} at \SI{200}{\celsius}.

\paragraph{Sample preparation}
Samples are prepared to
faithfully reflect the powder deposition at its most challenging position: in the centre of the printing volume.
The energy source for sintering (IR-lamp) provides
heat in a rather evenly distributed manner over an entire area,
enabling
solidification of
the entire sample-section within less than \SI{20}{\s}. 
However, due to design limitations the \gls{ir}-lamp provides slightly stronger heating on two regions of the printing bed of approx.~\SI{10}{\mm} by \SI{30}{\mm}, represented in blue in Fig.~\ref{fig:samples_photo}(a)--(b).
The lamp is placed accordingly to ensure that the centre of the printing bed be under one of the areas of preferred heating, as shown in Fig.~\ref{subfig:sample_heatzone}.
Each \gls{3d} printed sample is hence cut to extract a square sample of approx.~10 by \SI{10}{\mm}, cut out of the centre of the printing bed -- see Fig.~\ref{subfig:selecsample}.

\paragraph{Reference \enquote{known good} and \enquote{known bad} samples} 
One sample was sintered on-ground under the best possible conditions to serve as a reference \enquote{known good} sample for comparison with \gls{3d} printed ones. 
It is sintered from the highly flowable \gls{ss} powder in an oven at \SI{200}{\celsius} for one hour, under a weight of \SI{3}{\kg} to ensure continuous pressure \emph{during} sintering. This aims to increase material density, as the continuous pressure enhances degassing and porosity size reduction, akin to hot isostatic pressing \cite{Atkinson2000}. The resulting \enquote{known good} exemplary sample is shown in Fig.~\ref{subfig:known_good_xct}.

Another sample, \gls{3d} printed in \gls{mug} under wrongly tailored solidification parameters, serves as a \enquote{known bad} sample (see Fig.~\ref{subfig:known_bad_xct}).
In this sample, the solidification energy was too high, so that overheating resulted in partial melting instead of sintering into homogeneous layers.
As the empty space between particles becomes trapped in molten material,
the volume loss is not counteracted by reduction of the printing bed volume. Therefore, the supplementary volume transforms into large porosities scattered along the sample.
This sample is used here to validate the characterization procedure
by showing the results obtained for a \enquote{worst case scenario}.

\paragraph{\Acrshort{xct} specifications \& data analysis}
In-bulk characterization of the samples is done by \gls{xct}.
The machine and scanning parameters used are presented in Tab.~\ref{tab:TomoParameters}. 

\begin{table*}
	\scriptsize
    \centering
    \caption{\label{tab:TomoParameters} Scanning parameters for X-ray computed tomography of sintered \gls{ps}.}
	    \begin{tabular}[c]{C{0.16\textwidth}C{0.08\textwidth}C{0.07\textwidth}C{0.1\textwidth}C{0.13\textwidth}C{0.12\textwidth}C{0.09\textwidth}}
    \toprule
	\textbf{\Acrlong{ct} system}
	& \textbf{Source voltage}
	& {\textbf{Output current}}
	& \shortstack[c]{{\textbf{Projection}} \\ { \textbf{per scan}}}
	& {\textbf{Measurements per projection}}
	& {\textbf{Exposure time}}
	& {\textbf{Voxel size}} \\
	\midrule	
	\shortstack[c]{{CT-ALPHA} \\ {(ProCon X-ray, Germany)}} 
	& \shortstack{\SI{80}{\kilo\volt} \\ {}} 
	& \shortstack{\SI{70}{\micro\ampere} \\ {}} 
	& \shortstack{1600 \\ {}} 
	& \shortstack{10 \\ {}} 
	& \shortstack{\SI{1000}{\milli\second} \\ {}} 
	& \shortstack{{\SI{8}{\micro\meter}} \\ {}}  \\
    \bottomrule
    \end{tabular}  
\end{table*}

The size-homogeneity of the porosities in the samples is assessed
by two automatized image analysis procedures implemented in the Python 
PoreSpy library~\cite{Gostick2019}.
First, 
images are made binary by the automated procedure available in ImageJ~\cite{Schneider2012} (\textit{intermodes} method, automated thresholding).
Then, an average
density is calculated per \enquote{slice} of depth \SI{8}{\micro\meter} along the $y$-axis. 
The pore size distribution is found
by determining for each pore the maximal radius of a sphere that fits inside.
This method is adapted for samples showing relatively spherical porosities, homogeneous in shape, which is the case for most of our \gls{3d} printed samples. 
To capture the length of pores with \enquote{un-spherical} shapes 
and identify a possible anisotropy in pore shape,
the \enquote{chord length} method is employed. It consists 
of drawing chords that span across each pore in a given direction; the appearance frequency of each chord length is extracted along $x$- and $z$-directions and compared,
to detect 
large defects, in the deposition direction ($xy$-plane) or regarding interlayer adhesion ($z$-direction).

%% file: Results.tex
\begin{figure}[h!]
	\centering
    \subfigure[\Gls{rs} \gls{mug} sample \label{subfig:B2}]{
    \includegraphics[height=0.4\linewidth, trim={70pt 10pt 210pt 0pt}, clip]{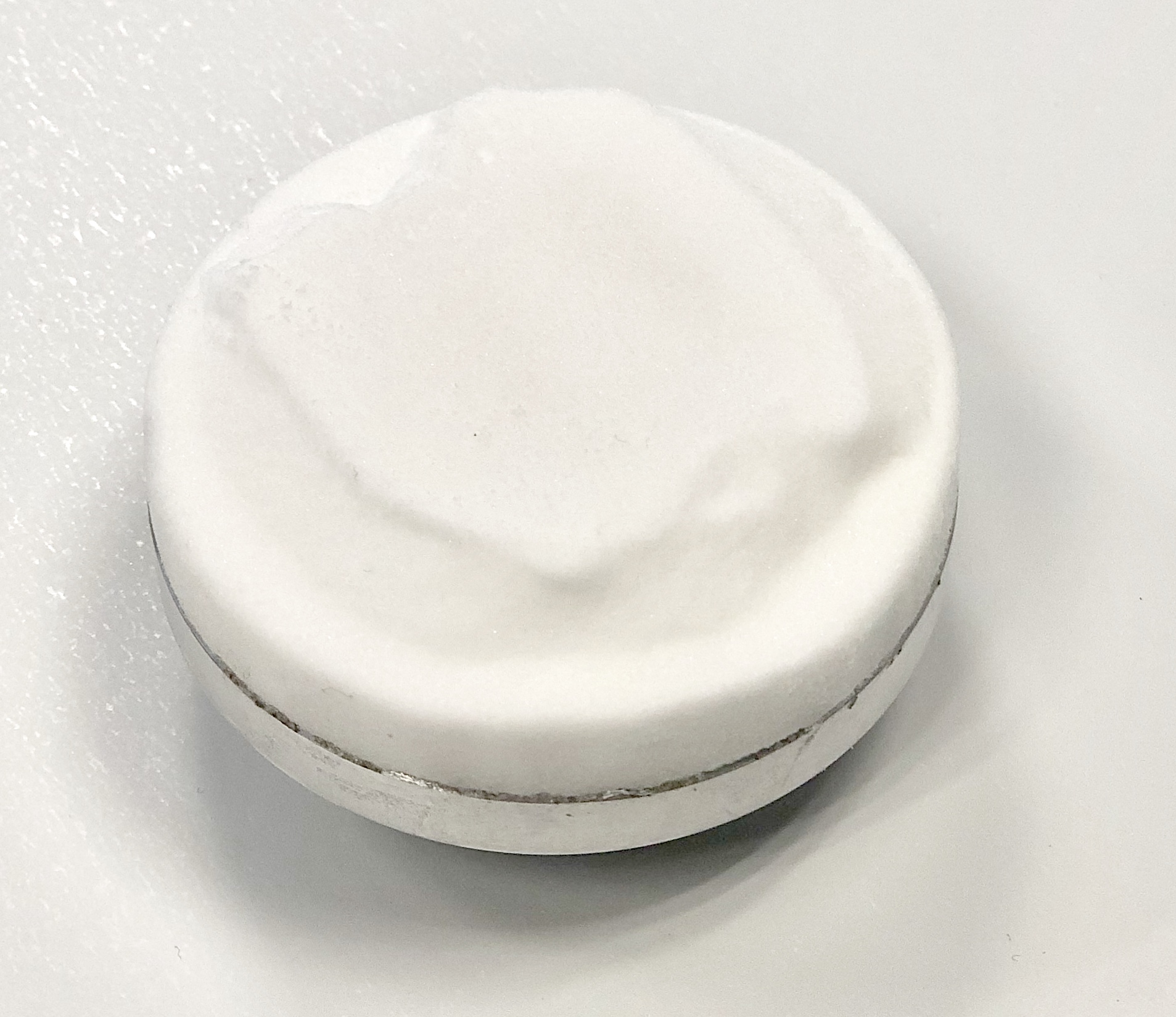}}
	\subfigure[\Gls{ss} \gls{mug} sample ]{
     \includegraphics[height=0.4\linewidth, trim={160pt 10pt 190pt 0pt}, clip]{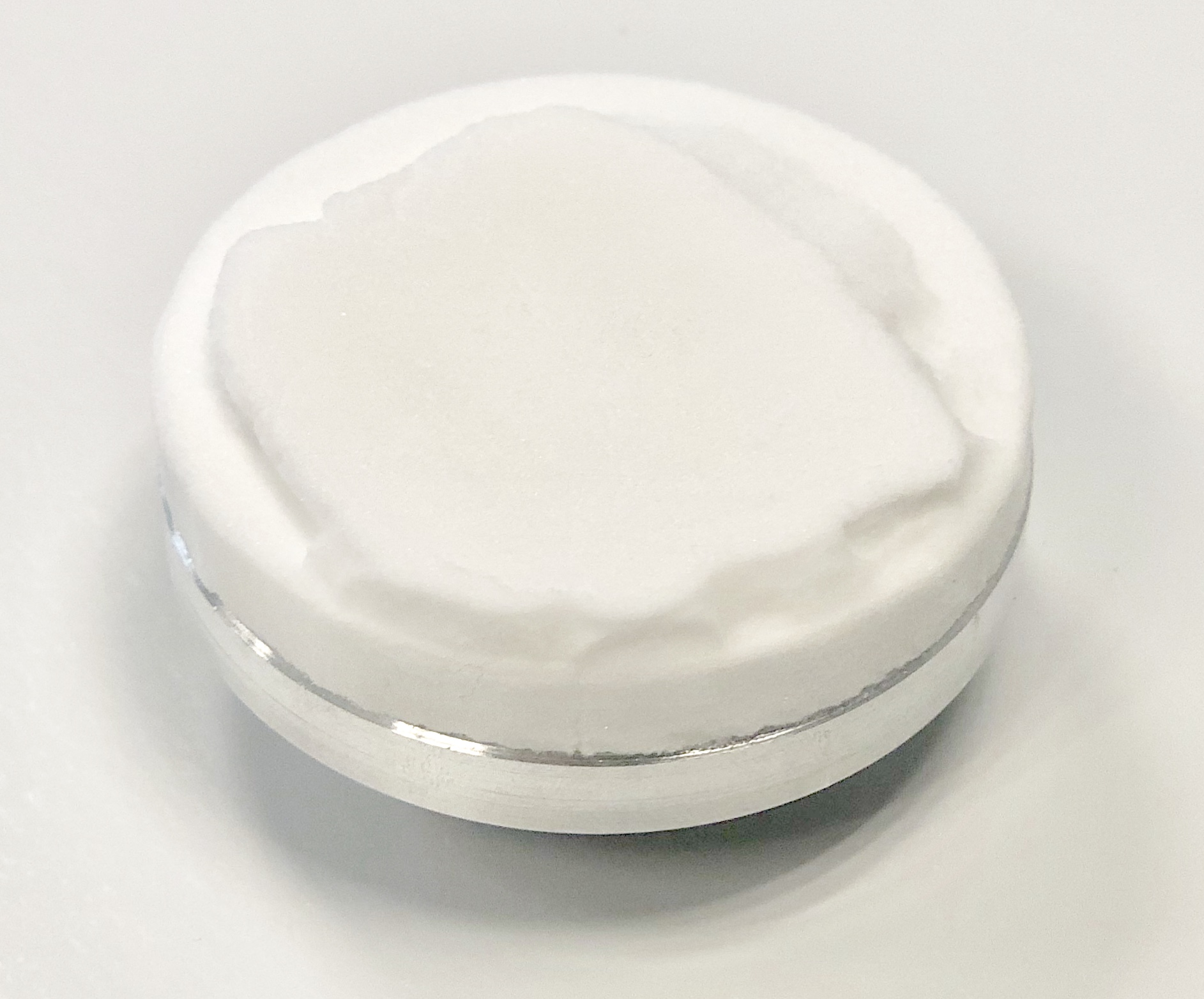}}
	\caption{\label{fig:samples} Samples \gls{3d} printed in \gls{mug}, from (a)~\gls{rs} and (b)~\gls{ss} powders, as extracted from the printing bed.}
\end{figure}

The primary result to report
is the successful manufacturing of samples \gls{3d} printed from 
the \gls{ss} and \gls{rs} powders,
under \gls{1g} and \gls{mug} (see Fig.~\ref{fig:samples}).
The resulting parts, 
constituted of up to 15 layers deposited successively in \gls{mug} and up to 20 under \gls{1g},
maintain their shape,
show homogeneous external appearance and smooth surface, without obvious defects, holes,
nor heterogeneous powder repartition.
Samples were put through further analysis to verify if this macroscopic assessment could be extended to the microscopic scale.

\begin{figure}[h!]
	\centering
	\subfigure[\label{subfig:layer_rising_before} Before platform rise]{
	\includegraphics[width=0.46\linewidth]{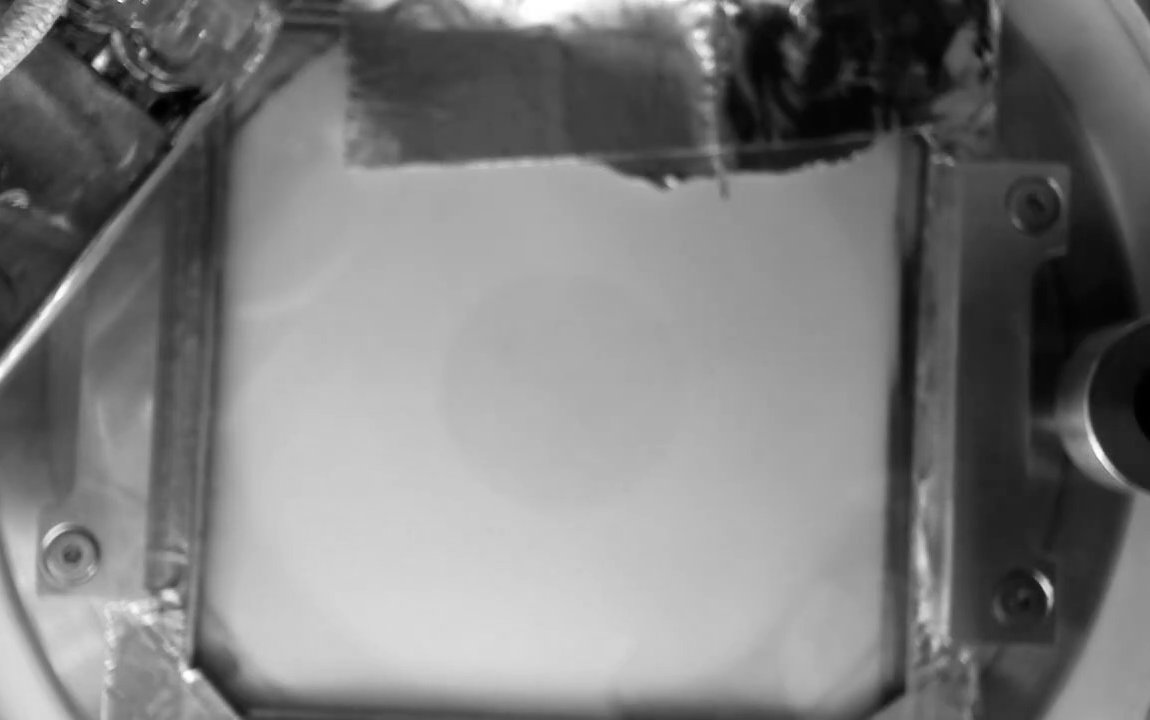}
	}
	\subfigure[\label{subfig:layer_rising_after} After platform rise]{
	\includegraphics[width=0.46\linewidth]{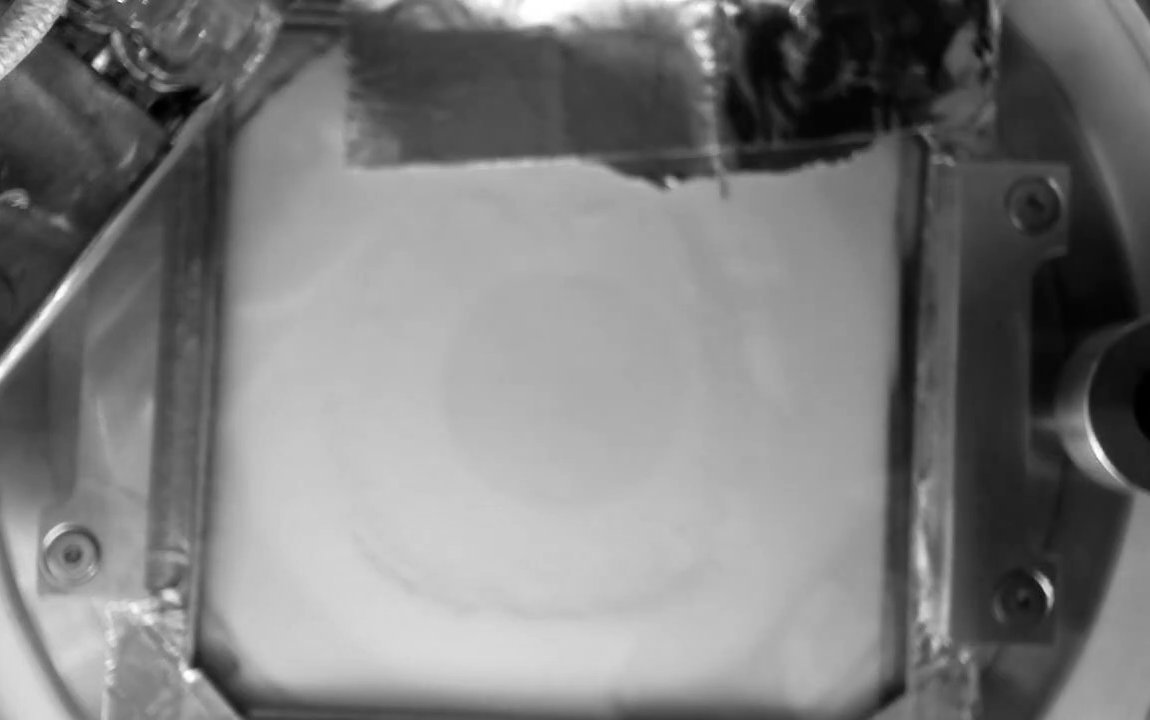}
	}\\
	\subfigure[\label{subfig:layer_rising_substraction} Difference between images (b) and (a) ]{
	\includegraphics[width=0.95\linewidth]{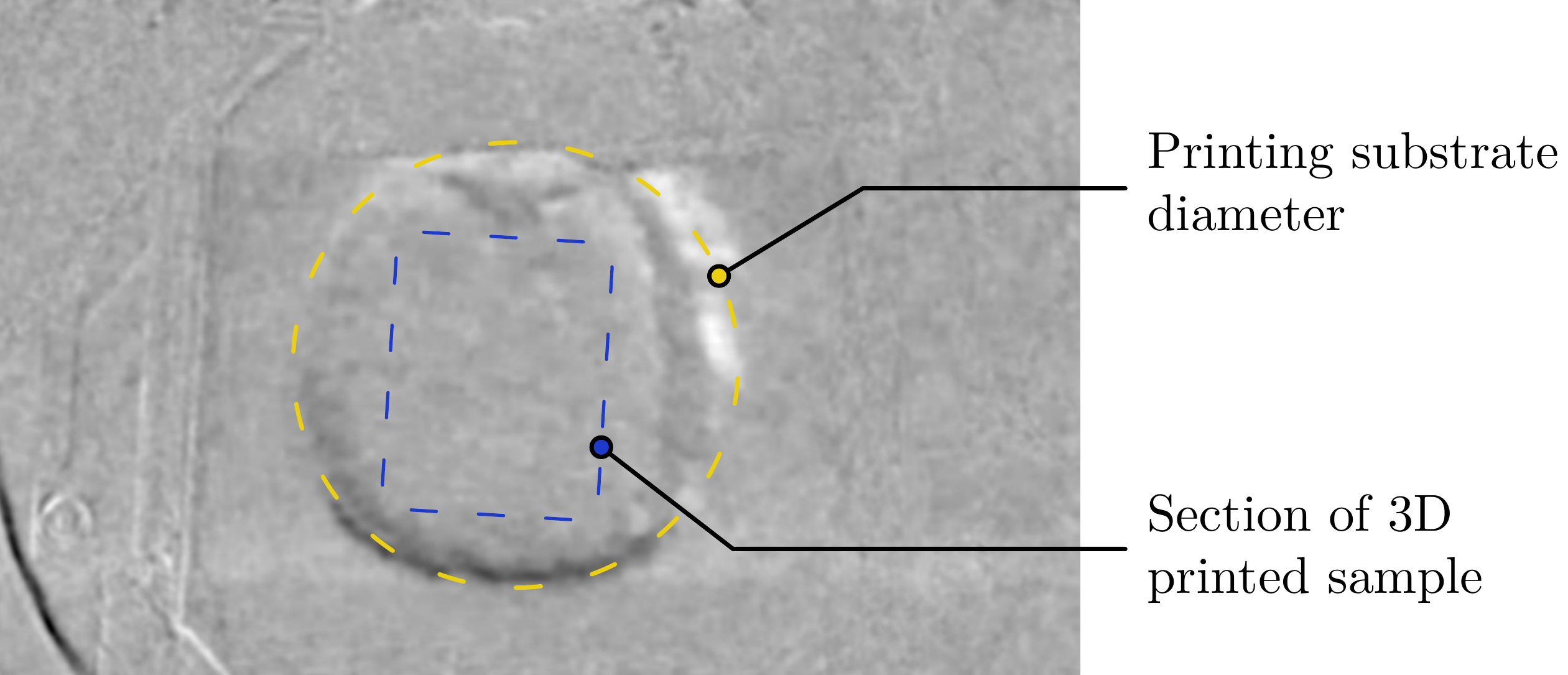}
	}
	\caption{\label{fig:layer_rising} 
Snapshots of \textit{in-situ} monitoring by image capture from below the solidification window.
The extract corresponds to a sample \gls{3d} printed from \gls{ss} base granular material, in \gls{mug} (third day of \gls{pfc}),
for the $4^{\text{th}}$ layer deposited
at time: (a)~just before the rise of the platform following the partial sintering of layer 3 and (b)~after platform rise. 
Panel (c)~shows the difference between (b) and (a);
the dashed yellow circle indicates the approximate boundary of the printing substrate,
and the dashed blue rectangle indicates the region corresponding to the previously
solidified part of the layer (i.e.~the zone where the lamp heating is strongest). 
}
\end{figure}

As a side remark, 
the images captured from under the printing area (below the solidification window)
show that at each rise of the printing substrate,
the entire consolidated layer (including its sintered and non-sintered regions) rises simultaneously, leaving the bottom space free for the next layer.
For printing complex shapes, 
it is important that
the homogeneity of the formerly deposited layers remains intact,
as it allows to distribute compressive stresses throughout the former layers (where \enquote{former layers} comprises the already printed sample \emph{and} the raw material situated in the non-solidified spaces).
In our test case,
where the \gls{3d} printed samples are of rectangular cross section and placed 
in the middle of a circular printing substrate,
the rise of the entire consolidated powder layer is shown in 
Fig.~\ref{fig:layer_rising}:
a difference image obtained from \textit{in-situ} imaging just
after (Fig.~\ref{subfig:layer_rising_after}) and just before (Fig.~\ref{subfig:layer_rising_before})
the rise of the inner cylinder, shows that the entire printing substrate
can be identified (Fig.~\ref{subfig:layer_rising_substraction}; yellow dashed circle),
but the previously solidified rectangular shape cannot be distinguished (dashed blue line).
The full video is available in supplementary material (video 2)~\cite{sm}. 
The fact that the layer remains consistent during and after the substrate's rise is attributed to the compression step, which consolidates the layer before solidification by sintering.
This conclusion emphasizes the importance of the compression step.

\subsection{Density of \gls{3d} printed samples}

\begin{figure}[h!]
	\centering
	\includegraphics[width=\linewidth]{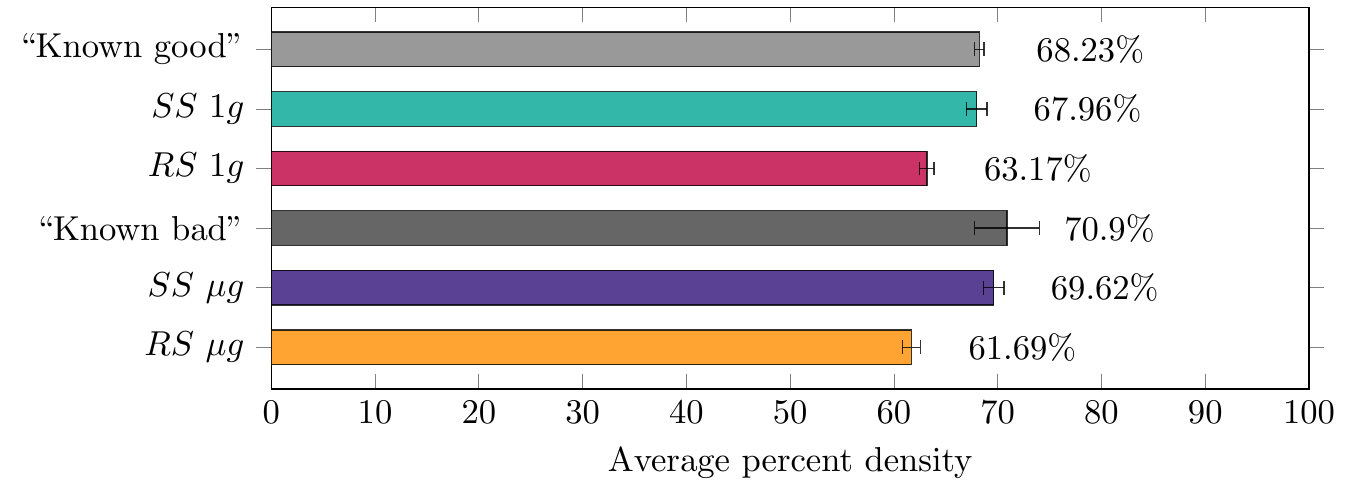}
	\caption{Percent area of material \textit{versus} voids per slice for each sample, representing the average density of the samples. Error bars represent the standard deviation per $xz$-slice percent area for all slices, in each sample.}
	\label{fig:densities}
\end{figure}

\begin{figure*}
\centering
\subfigure[\enquote{Known good}, \gls{ss} powder in \gls{1g} \label{subfig:known_good_xct}]{
\includegraphics[width=0.3\textwidth]{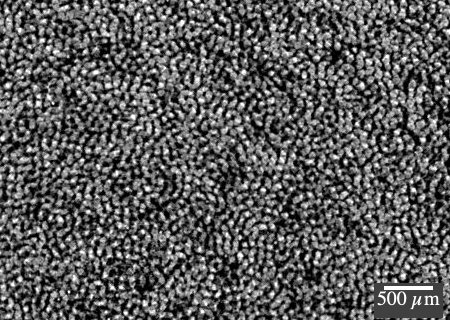}}
\subfigure[\gls{ss} powder in \gls{1g}]{
\includegraphics[width=0.3\textwidth]{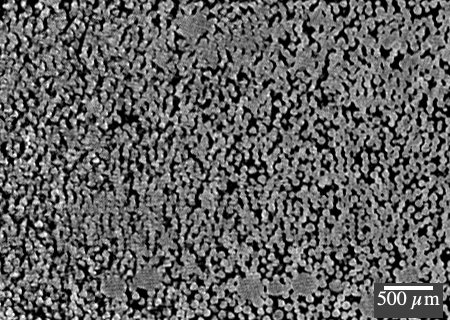}}
\subfigure[\gls{rs} powder in \gls{1g}]{
\includegraphics[width=0.3\textwidth]{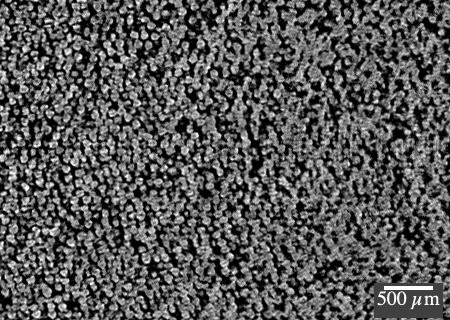}}
\subfigure[\enquote{Known bad}, \gls{rs} powder in \gls{mug} \label{subfig:known_bad_xct}]{
\includegraphics[width=0.3\textwidth]{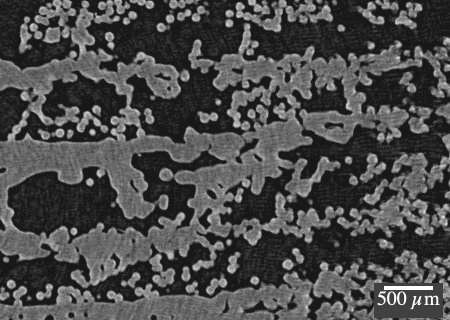}}
\subfigure[\gls{ss} powder in \gls{mug}]{
\includegraphics[width=0.3\textwidth]{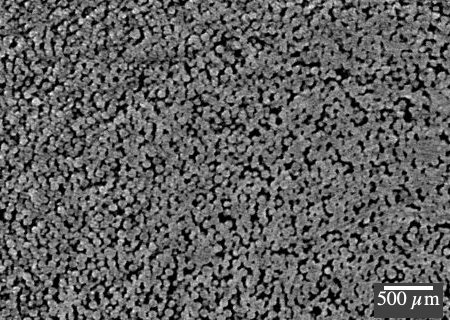}}
\subfigure[\gls{rs} powder in \gls{mug}]{
\includegraphics[width=0.3\textwidth]{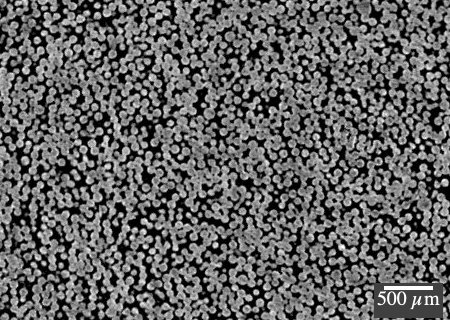}}
\caption{Extracts from \gls{xct} slices of samples \gls{3d} printed in all situations studied.
$xz$-plane is shown, $x$ to the right, $z$ to the top, with $z$ being the height (direction orthogonal to the layers). (a)~is the \enquote{known good} sample: \gls{ss} powder sintered under weight on-ground. (b)~ is the \gls{ss} powder and (c)~the \gls{rs} powder, both \gls{3d} printed in \gls{1g}. (d)~is the sample labeled \enquote{known bad} (partially molten \gls{rs} powder in \gls{mug}, with no compression during melting). (e)~and (f)~are respectively \gls{ss} and \gls{rs} powders \gls{3d} printed in \gls{mug}.
}
\label{fig:tomo_samples}
\end{figure*}

The average packing fraction
 of each sample is compared for all samples in Fig.~\ref{fig:densities}
by calculating the percent area of material \textit{versus} voids per slice, for each sample.
Since the slices have a finite thickness that is much smaller than the typical particle size, the average over the different slices thus represents the volume fraction of material in the sample.
This allows 
for verification of
the quantity of the material that was effectively deposited.
Typical \gls{xct} slices used for in-bulk characterization are shown in Fig.~\ref{fig:tomo_samples}.
Visual inspection of the \gls{xct} slices shows a rather homogeneous
distribution of the porosities throughout the sample, for all but the \enquote{known bad} sample. The spatial homogeneity is also confirmed by the fact that the distribution of average densities calculated per slice is very narrow (low standard deviation indicated in Fig.~\ref{fig:densities} and discussed below).
Bare-eye observation of the \gls{xct} images also shows that
all porosities display a relatively spherical shape, with neither a preferred direction nor obvious signs of delamination between layers,
for all samples but the \enquote{known bad} one.
Notably, the 
\SI{500}{\micro\meter}-high layers cannot be distinguished with bare eye, although the samples each comprise multiple layers.

As the samples are sintered and not melted, the average density of a good quality \gls{3d} printed part is expected to be slightly higher than \acrfull{rcp} for a monodispersed spheres packing, i.e.~$\approx64\%$~\cite{OHern2002, Bernal1960}. 
The average densities of the \gls{3d} printed samples shown in Fig.~\ref{fig:densities} are higher than \gls{rcp},
showing that the powder is effectively deposited in the centre of the printing bed.
The \enquote{known good} sample sintered under compression and the \gls{ss} samples \gls{3d} printed under \gls{1g} and \gls{mug} reach $\approx$~70\% density, with standard deviations less than 1\%. The \gls{rs} samples show a density $\approx$~5\% lower, regardless of $g$-level and of the fact that the same printing parameters are deployed for both types of feedstock material.
Although higher porosity is reflected in the lower average density of the \gls{rs} samples,
very low
standard deviations of 0.7\% and 0.9\% (respectively for \gls{1g} and \gls{mug})
show that
mediocre flowability powder can be used as \gls{am} base-material
without triggering major defects in the printed parts.
Comparatively, the \enquote{known bad} sample shows a surprisingly high average density of $\approx$~70\%, 
but also high variability,
with a standard deviation of approx.~3\% (more than three times higher than for all other samples),
as from one \enquote{slice} to the next the large porosities observed overtake most of the sample or are reduced to a minimal volume.
This shows that regardless of the sample quality (unquestionably bad because of heterogeneous, large porosities),
powder is indeed transported to the printing region:
although the solidification energy was overestimated, which produced an unevenly melted sample, the powder deposition procedure functioned and brought the necessary quantity of material to the printing bed.

The average {packing fractions} depend on the base material, but are independent of gravity.
This result is reminiscent of the printing-zone densities obtained through simulation (Figs.~\ref{fig:plot_gravity}, \ref{fig:distance_from_center}, \ref{fig:cohesion_simu}):
the observed changes are minimal between \gls{1g} and \gls{mug}, but the
\gls{rs} powder is more difficult to deposit.
Simulation predicted that the homogenization step would erase material-dependence, but experiments show that \gls{rs} powder remains at lower density after homogenization and compression.
This discrepancy might be due to simulated particles being smooth in essence,
leading to lower reliability of our model for particles with roughened surface.

\subsection{Influence of feedstock flowability on pore size}

\begin{figure}
    \centering
    \includegraphics[width=\linewidth, trim=0mm 0mm 3mm 0mm, clip]{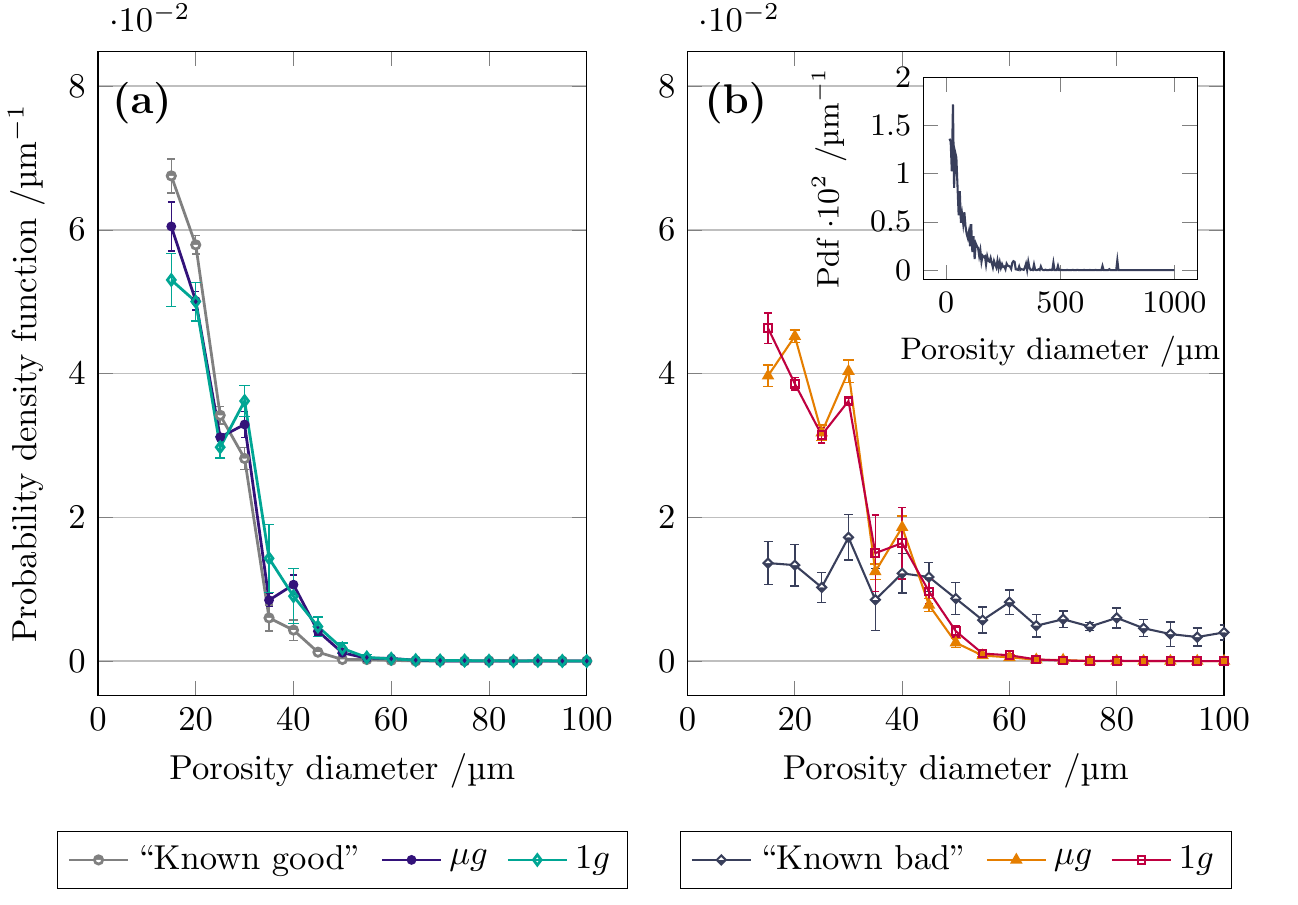}
    \caption{\label{fig:poresSizeDistrib_1g} \Acrfull{pdf} per circle diameter fitted in each pore for (a)~the \enquote{known good} sample, and the \gls{ss} samples \gls{3d} printed under \gls{mug} and \gls{1g}; (b)~the \enquote{known bad} sample, and \gls{rs} samples \gls{3d} printed under \gls{mug} and \gls{1g}.
    Points represent the average over each full sample, and error bars the standard deviation in one sample. The inset in (b) shows the \enquote{known bad} sample on appropriate scale.}
    \label{fig:porosities}
\end{figure}

The pore size distribution is shown in Fig.~\ref{fig:porosities}.
The \gls{xct} imaging resolution sets the 
minimum pore size detectable at \SI{15}{\micro\meter}.
The comparison between the \gls{ss} samples in Fig.~\ref{fig:porosities}(a) shows that
all have most pores in the range of \SIrange{15}{25}{\micro\meter}.
The sample sintered under weight (\enquote{known good} sample) has the largest amount of small pores (diameter \SI{15}{\micro\meter}), 
which shows, as expected, the best powder repartition and compaction.
The \gls{3d} printed samples also show a peak at small pores;
however for both material-qualities, a second peak appears at \SI{30}{\micro\meter}, increasing the mean pore size.
Most importantly, 
large pores of \SIrange{35}{60}{\micro\meter} have a very low probability in the \enquote{known good} sample, and low for the \gls{ss} \gls{3d} printed samples.
Despite small differences, the three samples have mostly similar pore size distributions, and
consistently show virtually
no large pores of size~$\geq$~\SI{55}{\micro\meter}.

The \gls{rs} samples in Fig.~\ref{fig:porosities}(b) show an obvious difference in
pores size for the \enquote{known bad} sample:
its pore diameter distribution has a much longer tail than all other samples, with a low probability density of small pores under \SI{50}{\micro\meter} and some very large pores up to \SI{700}{\micro\meter}.
Apart from this outlier experiment, our \gls{rs} results (Fig.~\ref{fig:porosities}b) in comparison to the \gls{ss} results (Fig.~\ref{fig:porosities}a) confirm that a clear
difference between base materials can be made: even if the distribution is relatively close, with two peaks at pore diameters \SI{15}{\micro\meter} and \SI{30}{\micro\meter}, for the smooth-surface samples 
the smaller pore size is almost twice more likely to appear than the smaller one, while for the \gls{rs}, the two pore diameters are almost equally likely to appear. 
Figure~\ref{fig:porosities also} confirms that the samples printed on-ground and in \gls{mug} from their respective base-materials show very similar pore size distributions: within experimental error, the gravitational environment in which the samples have been manufactured does not play a role in the quality of \gls{3d} printed samples.

The pore size distribution corresponding to the \gls{rs} powder has a larger width than that corresponding to the \gls{ss} powder (Fig.~\ref{fig:porosities}). Hence, the \gls{rs} sample has a higher fraction of larger pores 
than the \gls{ss} sample, showing that 
its lower average density (Fig.~\ref{fig:densities})
is the result of generally larger pores, similarly well scattered throughout the samples (from visual inspection of the \gls{xct} slices).
Such larger porosities are linked to the decreased packing efficiency of materials exhibiting increased surface roughness: 
as the stress needed for particles to slide on each other is increased by surface roughness \cite{DeGiuli2016}, 
under the same external stress input,
the packing reorganizes into a less dense configuration~\cite{Mari2014}.
The maximum packing density for each specific type of powder is effectively reached in our experiments, in absence of other means implemented to increase material density (e.g.~compression during sintering).

\subsection{Gravity-dependent anisotropy}

\begin{figure*}[h!]
    \centering 
    	\includegraphics[width=0.8\linewidth]{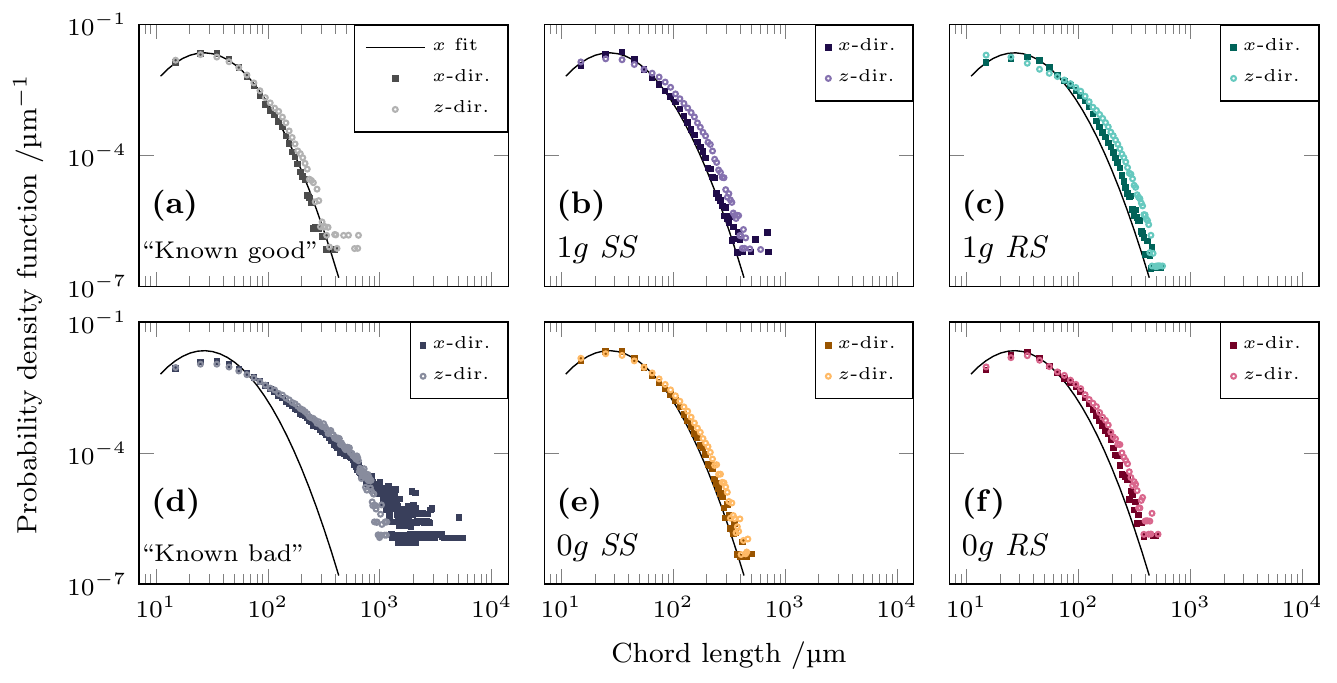}
    \caption{Relative frequency of chord spanning porosities per chord length, for all samples studied, with chords extension done along $x$- and $z$-directions for each sample. The \enquote{known good} sample data along the $x$-axis is fitted to a log-normal distribution (solid black line), reproduced on all graphs for comparison.}
     \label{fig:chord_length_full} 
\end{figure*}

Chord length analysis is used
to capture anisotropy in the pores' shape and test whether the manufacturing process begets a preferred direction; 
it is presented in
Fig.~\ref{fig:chord_length_full} for the $x$- and $z$-directions.
To allow precise comparison of the different samples, each data set is fitted with a log-normal distribution. 
This distribution is plausible because the chord lengths are not independent: each pore that contributes to chord length $L$, also
contributes to all smaller chord lengths, so that the addition of a long chord rescales the entire distribution at smaller chord lengths. Assuming that pore sizes are independent from each other and randomly distributed, the central limit theorem hence suggests a log-normal distribution (since the multiplicative increments become independent random additive increments of the logarithm).
Previous work involving the
distribution of geometrical shapes enclosed within randomly distributed voids
also found a log-normal distribution~\cite{Marakis2019}.
The fitting parameters, $\mu_{x, \,z}$ and $\sigma_{x, \,z}$, corresponding to the mean and standard deviation of the logarithm of the data, are presented for each sample in $x$- and $z$-directions in Tab.~\ref{tab:fittingparametersChordLength}. The mean chord length $\langle l_{x, \, z} \rangle = \exp{(\mu_{x, \, z} +\nicefrac{\sigma_{x, \, z}^2}{2})} $ is also given for each fit to ease interpretation. 

\begin{table*}[h!]
    \centering
    \footnotesize
    \caption{\label{tab:fittingparametersChordLength} Fitting parameters used to fit the chord length probability density obtained for each sample in $x$- and $y$-directions to the log-normal distribution, $\mu_{x, \, z}$ and $\sigma_{x, \, z}$, and mean chord length $\langle l_{x, \, z} \rangle$, given in~\si{\micro\meter}. The ratio of standard deviations, $\nicefrac{\sigma_z}{\sigma_x}$, is also given for reference.}
    \renewcommand*{\arraystretch}{1.2}
    \begin{tabular}[c]{L{0.2\textwidth}C{0.04\textwidth}C{0.04\textwidth}C{0.04\textwidth}C{0.04\textwidth}C{0.04\textwidth}C{0.04\textwidth}C{0.04\textwidth}}
    \toprule
    {Sample reference} 
    & {$\mu_x$} & {$\sigma_x$}  & {$\langle l_{x} \rangle$} & {$\mu_z$} & {$\sigma_z$} & {$\langle l_{z} \rangle$} & {$\nicefrac{\sigma_z}{\sigma_x}$}\\ 
    \midrule
	{\enquote{Known good} \quad} & {3.6} & {0.57} & {44} & {3.7} & {0.63} & {48} & {1.105} \\ 
	Smooth surface, \gls{1g} & {3.7} & {0.57} & {48} & {3.8} & {0.65} & {57} & {1.140} \\
	Rough surface, \gls{1g} & {3.9} & {0.54}  & {56} & {3.9} & {0.67} & {60} & {1.241} \\
	\enquote{Known bad} & {3.9} & {1.1} & {92} & {4.1} & {1.1} & {110} & {1} \\
	Smooth surface, \gls{mug} & {3.7} & {0.58} & {47} & {3.7} & {0.64} & {51} & {1.103} \\
	Rough surface, \gls{mug} & {3.8} & {0.59} & {56} & {3.9} & {0.67} & {60} & {1.136} \\
    \bottomrule
    \end{tabular} 
\end{table*}

For the \enquote{known good} sample in Fig.~\ref{fig:chord_length_full}(a), without surprise
the mean chord lengths are the lowest of all experiments.
They are also very close in size along $x$- and $z$-directions, with $\langle l_{z} \rangle$ larger than $\langle l_{x} \rangle$ by only \SI{4}{\micro\meter}.
The fit for this dataset ($x$-dir.) is shown on all panels of Fig.~\ref{fig:chord_length_full} (solid black line) to provide a comparative baseline.

Comparison of the \gls{3d} printed samples to this baseline confirms
that all samples have slightly larger pores than our reference sample. 
The comparative sample is sintered continuously under weight, applying a constant pressure to allow porosities to close during sintering, while the \gls{3d} printed samples are all 
compressed to a fixed height rather than under a constant pressure:
the compression is not maintained constant during sintering. 
This is necessary for assessing the powder repartition, but a constant compression pressure should be maintained during the solidification step in further manufacturing campaigns to decrease the size of porosities, thereby increasing the prints' quality.

The \enquote{known bad} sample's chord length presented in Fig.~\ref{fig:chord_length_full}(d) is notable.
The distribution shows a long tail at large chord lengths $l_{x, \, y} \geq 10^3$~\si{\micro\meter} in $x$-direction, which does not appear along $z$-direction, 
indicating the long horizontal porosities due to delamination between layers, clearly visible in Fig.~\ref{fig:tomo_samples}(d).
This shows an exemplary anisotropic sample with preferential direction of pore growth along the $x$-direction, populated by many elongated porosities of length $l_{x} \geq 10^3$~\si{\micro\meter}.

The \gls{mug} \gls{ss} sample has the highest isotropy of all \gls{3d} printed samples.
The \gls{mug} \gls{rs} sample also exhibits high isotropy, despite
slightly larger porosities.
Generally, the raw material's flowability (\gls{ss} or \gls{rs}) does not beget 
anisotropy in the printed samples.

Comparing the \gls{3d} printed samples by base-material depending on the $g$-level during manufacturing, 
a slight elongation of the pores in the $z$-direction is remarked for the \gls{1g} samples.
Precisely, $\nicefrac{\sigma_z}{\sigma_x}$ is always larger for the \gls{1g} samples than for the corresponding \gls{mug} samples (see Table~\ref{tab:fittingparametersChordLength}), showing that \gls{1g} samples
have a stronger anisotropy than \gls{mug} samples.
To understand the origin if this divergence, 
we look at
the homogenization step of the \gls{3d} printing process.

In microgravity, the homogenization step
(horizontal shaking of the powder at the bottom of the apparatus),
results in a wave of powder forming \enquote{clusters} or zones of heterogeneous density: bubble-like circular structures progressing towards the center of the printing area.
The homogenization step of the process consists of merging those \enquote{powder clusters} into a homogeneous powder layer. 
This can be seen in video 2 of the supplementary material~\cite{sm}.
On-ground, the same wave-like progression is observed, although the circular structures within the powder are not observed.

In absence of the preferred acceleration direction due to gravitational acceleration, the \enquote{clusters} that form
are composed of particles
in configurations that also do not have a preferred direction,
which results in an isotropic packing.
Those clusters then merge into a powder layer; as this still happens in absence of gravity, the particles are not reordered in a denser packing due to their respective weight, but retain the isotropic configuration they had in the clusters.
In turn, this results in porosities with isotropic shapes for the microgravity-\gls{3d} printed samples, 
while the ground-\gls{3d} printed samples have reorganized under their own weight into a less isotropic packing.

\begin{figure}[h!]
	\subfigure[\label{subfig:microRS1g} \gls{rs} \gls{1g}]{
	\includegraphics[width=0.46\linewidth]{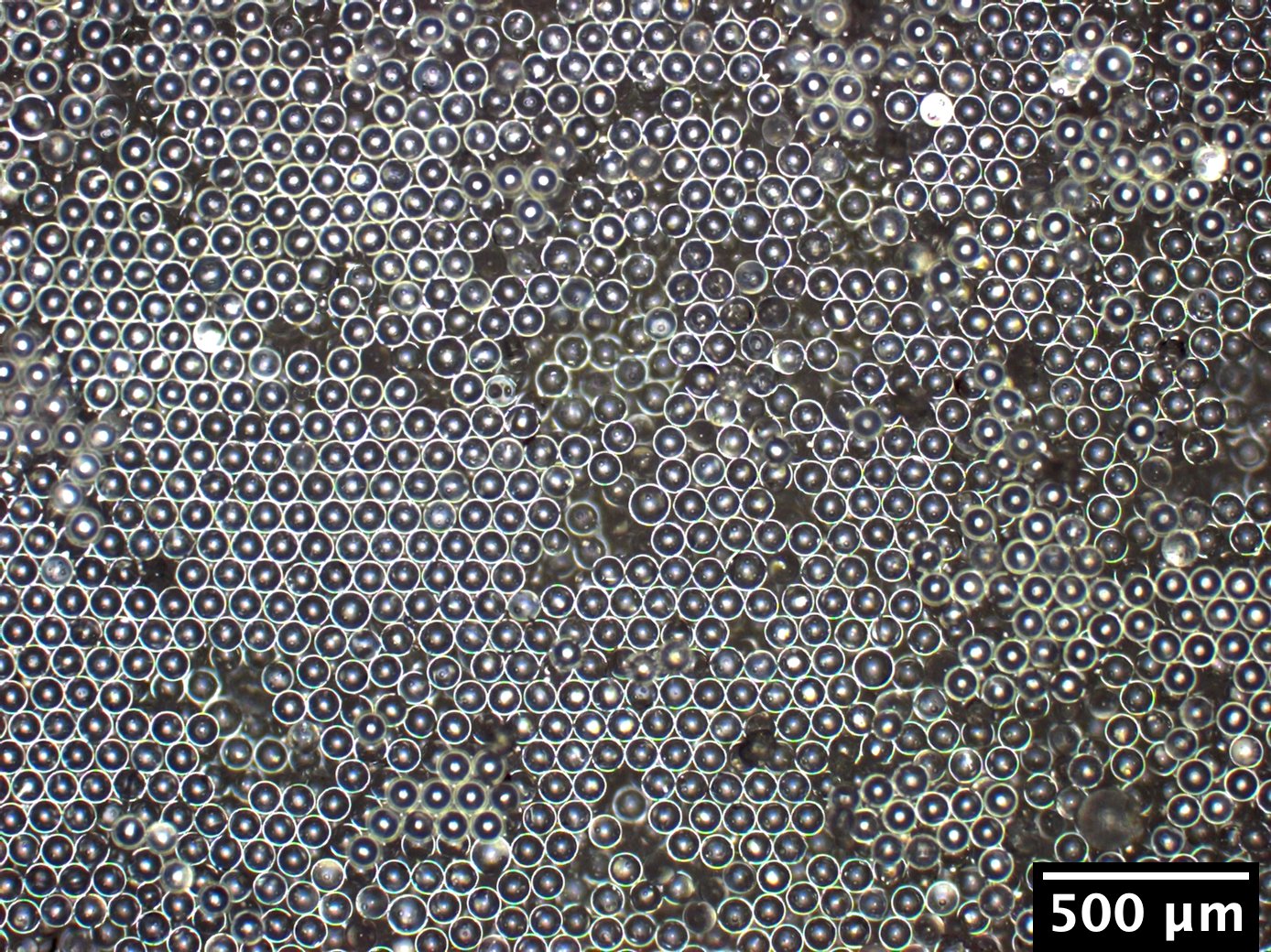}
	}
	\subfigure[\label{subfig:microRSmug} \gls{rs} \gls{mug}]{
	\includegraphics[width=0.46\linewidth]{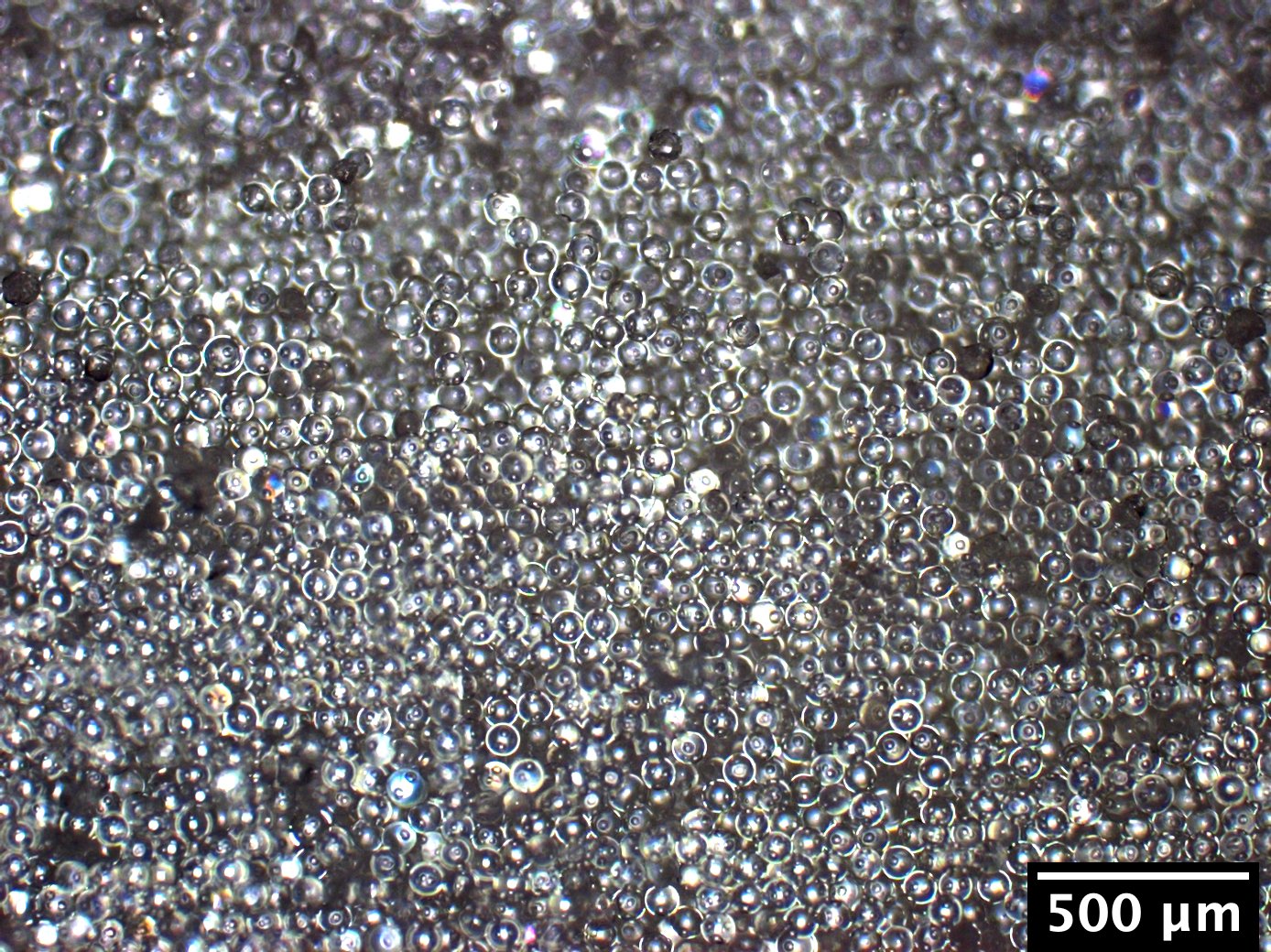}
	}
	\caption{\label{fig:micro_section} 
Bright field optical microscopies of sections cut of the \gls{3d} printed samples along the $xy$-direction, for samples manufactured from \gls{rs} powder, (a)~under \gls{1g} and (b)~in \gls{mug}.
}
\end{figure}

Microscopies of cuts of \gls{rs} samples that were \gls{3d} printed under \gls{1g} and \gls{mug} respectively, are presented in Figs.~\ref{subfig:microRS1g} and \ref{subfig:microRSmug}.
Although partial crystallization is visible on both samples, on the sample that was \gls{3d} printed under \gls{1g}, large crystallized areas are visible.
For both samples, the crystalline ordering is observed in the $xy$-plane, in which the horizontal shaking takes place.

Under \gls{1g} particles tend to crystallize layer-wise in the $xy$-plane.
Such self-ordering has been shown previously~\cite{Pouliquen1997}; it results in a superposition of high-density, crystallized grain-like regions, surrounded by lower density boundaries (as visible in Fig.~\ref{subfig:microRS1g}). As this phenomenon occurs along the shaking direction in the $xy$-plane, it creates voids elongated in the $z$-direction,
and explains the mild anisotropy
observed in samples manufactured under \gls{1g}.

The
minute elongation of porosities along the $z$-axis in the ground-manufactured samples
is generally compliant with the presence of a preferred direction under gravity,
and partial crystallization is observed in the horizontal shaking plane, 
in particular in the \gls{1g} samples (Fig.~\ref{fig:micro_section}).
The pores' chord length distributions are very similar to the comparative \enquote{known good} samples: all \gls{3d} printed samples show high isotropy (Fig.~\ref{fig:chord_length_full} and Tab.~\ref{tab:fittingparametersChordLength}). 
The decrease in density of the \gls{rs} samples is linked to an increase in pore size, but not accompanied by the appearance of defects in the deposition.
The quantity of material brought to the printing area, as well as the homogenization time (representing the amount of material brought down and then to the printing-bed centre) are hence deemed sufficient.